\documentclass[usenatbib]{mnras}

\usepackage{xspace}
\usepackage{xfrac}
\usepackage[table]{xcolor}
\usepackage{orcidlink}

\newcommand{\lya}{Lyman-$\alpha$\xspace}
\newcommand{\lyaf}{Lyman-$\alpha$ forest\xspace}
\newcommand{\varlss}{\sigma^2_{\rm LSS}}
\newcommand{\lambdarf}{\lambda_{\rm RF}}

\title[The Ly-$\alpha$ catalog from the DESI EDR]{The \lyaf catalog from the Dark Energy Spectroscopic Instrument Early Data Release}
\author[]{
	\parbox{\textwidth}{
		\Large
		C\'esar\ Ram\'irez-P\'erez,$^{1}$\orcidlink{0000-0003-1258-9353}
		Ignasi\ P\'erez-R\`afols,$^{2}$\orcidlink{0000-0001-6979-0125}
		Andreu\ Font-Ribera,$^{1,3}$\orcidlink{0000-0002-3033-7312}
		M.~Abdul Karim,$^{4}$\orcidlink{0009-0000-7133-142X}
		E.~Armengaud,$^{4}$\orcidlink{0000-0001-7600-5148}
		J.~Bautista,$^{5}$
		S.~F.~Beltran,$^{6}$\orcidlink{0000-0001-6324-4019}
		L.~Cabayol-Garcia,$^{1}$
		Z.~Cai,$^{7,8,9}$\orcidlink{0000-0001-8467-6478}
		S.~Chabanier,$^{10}$\orcidlink{0000-0002-5692-5243}
		E.~Chaussidon,$^{10}$\orcidlink{0000-0001-8996-4874}
		J.~Chaves-Montero,$^{1}$\orcidlink{0000-0002-9553-4261}
		A.~Cuceu,$^{11,12,13}$\orcidlink{0000-0002-2169-0595}
		R.~de la Cruz,$^{6}$\orcidlink{0000-0001-9908-9129}
		J.~Garc\'ia-Bellido,$^{14}$\orcidlink{0000-0002-9370-8360}
		A.~X.~Gonzalez-Morales,$^{15,6}$\orcidlink{0000-0003-4089-6924}
		C.~Gordon,$^{1}$
		H.~K.~Herrera-Alcantar,$^{6}$\orcidlink{0000-0002-9136-9609}
		V.~Ir\v{s}i\v{c},$^{16}$\orcidlink{0000-0002-5445-461X}
		M.~Ishak,$^{17}$\orcidlink{0000-0002-6024-466X}
		N.~G.~Kara{\c c}ayl{\i},$^{11,18,12,13}$\orcidlink{0000-0001-7336-8912}
		Zarija Luki\'c,$^{10}$
		C.~J.~Manser,$^{19,20}$\orcidlink{0000-0003-1543-5405}
		P.~Montero-Camacho,$^{9}$\orcidlink{0000-0002-6998-6678}
		L.~Napolitano,$^{21}$\orcidlink{0000-0002-5166-8671}
		G.~Niz,$^{6,22}$\orcidlink{0000-0002-1544-8946}
		M.M.~Pieri,$^{23}$
		C.~Ravoux,$^{5,4}$\orcidlink{0000-0002-3500-6635}
		F.~Sinigaglia,$^{24}$\orcidlink{0000-0002-0639-8043}
		T.~Tan,$^{25}$
		M.~Walther,$^{26,27}$\orcidlink{0000-0002-1748-3745}
		B.~Wang,$^{28,9}$\orcidlink{0000-0003-4877-1659}
		J.~Aguilar,$^{10}$
		S.~Ahlen,$^{29}$\orcidlink{0000-0001-6098-7247}
		S.~Bailey,$^{10}$\orcidlink{0000-0003-4162-6619}
		D.~Brooks,$^{30}$
		T.~Claybaugh,$^{10}$
		K.~Dawson,$^{31}$
		A.~de la Macorra,$^{32}$\orcidlink{0000-0002-1769-1640}
		G.~Dhungana,$^{33}$\orcidlink{0000-0002-5402-1216}
		P.~Doel,$^{30}$
		K.~Fanning,$^{13}$\orcidlink{0000-0003-2371-3356}
		J.~E.~Forero-Romero,$^{34,35}$\orcidlink{0000-0002-2890-3725}
		S.~Gontcho A Gontcho,$^{10}$\orcidlink{0000-0003-3142-233X}
		J.~Guy,$^{10}$
		K.~Honscheid,$^{11,12,13}$
		R.~Kehoe,$^{33}$
		T.~Kisner,$^{10}$\orcidlink{0000-0003-3510-7134}
		M.~Landriau,$^{10}$\orcidlink{0000-0003-1838-8528}
		L.~Le~Guillou,$^{25}$\orcidlink{0000-0001-7178-8868}
		Michael~E.~Levi,$^{10}$\orcidlink{0000-0003-1887-1018}
		C.~Magneville,$^{4}$
		P.~Martini,$^{11,18,13}$\orcidlink{0000-0002-4279-4182}
		A.~Meisner,$^{36}$\orcidlink{0000-0002-1125-7384}
		R.~Miquel,$^{37,1}$
		J.~Moustakas,$^{38}$\orcidlink{0000-0002-2733-4559}
		E.~Mueller,$^{39}$
		A.~Muñoz-Gutiérrez,$^{32}$
		J.~Nie,$^{40}$\orcidlink{0000-0001-6590-8122}
		N.~Palanque-Delabrouille,$^{4,10}$\orcidlink{0000-0003-3188-784X}
		W.~J.~Percival,$^{41,42,43}$\orcidlink{0000-0002-0644-5727}
		G.~Rossi,$^{44}$
		E.~Sanchez,$^{45}$\orcidlink{0000-0002-9646-8198}
		E.~F.~Schlafly,$^{46}$\orcidlink{0000-0002-3569-7421}
		D.~Schlegel,$^{10}$
		H.~Seo,$^{47}$\orcidlink{0000-0002-6588-3508}
		G.~Tarl\'{e},$^{48}$\orcidlink{0000-0003-1704-0781}
		B.~A.~Weaver,$^{36}$
		C.~Yèche,$^{4}$\orcidlink{0000-0001-5146-8533}
		and Z.~Zhou$^{40}$
	}
	\vspace{0.4cm}
	\\
	\parbox{\textwidth}{
		Affiliations are listed at the end of the paper	
	}}

\date{Accepted XXX. Received YYY; in original form ZZZ}
\pubyear{2023}

\begin{document}
\label{firstpage}
\pagerange{\pageref{firstpage}--\pageref{lastpage}}
\maketitle

\begin{abstract}
We present and validate the catalog of \lyaf fluctuations for 3D analyses using the Early Data Release (EDR) from the Dark Energy Spectroscopic Instrument (DESI) survey. 
We used 88,511 quasars collected from DESI Survey Validation (SV) data and the first two months of the main survey (M2). 
We present several improvements to the method used to extract the \lya absorption fluctuations performed in previous analyses from the Sloan Digital Sky Survey (SDSS). In particular, we modify the weighting scheme and show that it can improve the precision of the correlation function measurement by more than 20\%. 
This catalog can be downloaded from \href{https://data.desi.lbl.gov/public/edr/vac/edr/lya/fuji/v0.3}{https://data.desi.lbl.gov/public/edr/vac/edr/lya/fuji/v0.3}, and it will be used in the near future for the first DESI measurements of the 3D correlations in the \lyaf.
\end{abstract}

\begin{keywords}
	dark energy -- large-scale structure of Universe -- intergalactic medium  -- catalogues
\end{keywords}

\section{Introduction}
\label{sec:introduction}
The \lyaf is a pattern of absorption features caused by neutral hydrogen in the Intergalactic Medium (IGM), typically observed in the spectra of distant quasars ($z>2$). Measurements of 1D correlations in a handful of high-resolution quasar spectra emerged as a powerful tool to study the large-scale distribution of matter \citep{Croft_1998,McDonald_2000}, opening a new field in the analysis of the high redshift universe and helping to constrain cosmological parameters.

Using data from the Baryon Oscillation Spectroscopic Survey (BOSS, \citealt{Dawson_2013}), the three-dimensional correlation function of absorption in the \lyaf was measured for the first time in \cite{Slosar_2011}. Shortly after that, the first measurement of the Baryonic Acoustic Oscillations (BAO) peak in the \lyaf was presented \citep{Busca_2013,Slosar_2013,Kirkby_2013}, using data from BOSS DR9 \citep{Lee_boss_dr9}.
These were followed by other BAO analyses using increasingly larger \lyaf datasets from BOSS \citep{Delubac_2015, Bautista_2017} and from the extended Baryon Oscillation Spectroscopic Survey (eBOSS, \citealt{Dawson_2016}) \citep{Agathe_2019}.
The precision of these BAO measurements was significantly improved with the measurement of the cross-correlation of quasars and the \lyaf \citep{Font-Ribera_2014, dMdB_2017, Blomqvist_2019}, and the final \lya BAO measurement combining BOSS and eBOSS was presented in \citet{dMdB_2020}. 

The Dark Energy Spectroscopic Instrument (DESI) is currently undergoing a five-year campaign to obtain close to a million quasar spectra with $z > 2$ \citep{sv_kp, qso_ts}. This dataset will be four times larger than the state-of-the-art (eBOSS DR16 quasar sample \citealt{Lyke_eboss_quasar_catalog}), and will enable sub-percent BAO measurements with the \lyaf \citep{snowmass_2013, desi_2016a}.

In this publication we present the first catalog of \lyaf fluctuations in DESI,
including data from the Early Data Release (EDR, \cite{edr_kp}) and from the first two months of the main survey (M2).
This dataset is used in a companion paper to measure the first 3D correlations
in the \lyaf from DESI \citep{edrmain}, and a comparison with
synthetic datasets is presented in \citet{edrmocks}.

The methodology used here is similar to the one developed for eBOSS analyses, especially the most recent analysis by \citet{dMdB_2020}. 
This served as the basis for developing the data analysis pipeline of the DESI \lyaf working group. 
In this publication, we provide a detailed description of our new pipeline, focusing on the changes with respect to the one used in eBOSS analyses. 
Some of these changes are motivated by changes in the input data: for instance, while SDSS spectra had pixels equispaced in the logarithm of the wavelength, DESI uses linearly spaced pixels. 
Other changes are motivated by studies that appeared after \citet{dMdB_2020}.
For instance, following \citet{BALsEnnesser} we now include in our analysis the spectra of Broad Absorption Line (BAL) quasars, after we mask the most contaminated regions. 
Finally, we also revisit the weighting scheme used to compute correlations in the \lyaf, resulting in an improvement of more than 20\% in our precision.

All of the process followed in this paper, along with the updated changes, is executed using the publicly available code \textsc{picca}\footnote{https://github.com/igmhub/picca/} and can be reproduced by the user using public DESI data. The \textsc{picca} package also includes modules for the computations of both auto- and cross-correlation with quasars, cosmological fits, and multiple useful tools for \lyaf studies.

The catalog described here is aimed at studies of 3D correlations in the \lyaf.
A similar dataset, however, is also used in two companion publications that present the first measurements of 1D correlations with DESI data
\citep{edr1d,edr1doptimal}.
Although analogous estimations of the unabsorbed quasar continuum are also needed in these studies, their methodology is somewhat different and these publications focus on studies of systematics that primarily affect the 1D correlations.

This paper is structured as follows. In Section \ref{sec:data}, we present the data used for this work, including spectra and quasar catalogs.
In Section \ref{sec:continuumfittingprocedure}, we explain how we obtain the flux-transmission field from spectra through the estimation of the expected flux of quasars in the continuum fitting process. 
This includes masking some wavelengths and applying corrections to the flux calibration and the reported uncertainties by the pipeline. 
In Section \ref{sec:discussion}, we discuss aspects that require further clarification, such as modified weights and wavelength grid choices. Finally, in Section \ref{sec:summary}, we provide a summary of our findings and conclusions.

\section{Data}
\label{sec:data}
Data used in this publication comes from two different DESI datasets. On the one hand, we have the Early Data Release (EDR), which includes all Commissioning, Survey Validation (SV), and special survey data. On the other hand, we have EDR+M2, which includes all data from EDR as well as the first two months of the main survey.

Although this two samples are qualitatively similar, we are performing the analysis separately for each of them to achieve the two purposes of this paper:
\begin{itemize}
	\item \textbf{Describe the \lyaf Value Added Catalog (VAC) in the context of EDR \citep{edr_kp}:} VACs for a wide variety of tracers are being released using DESI early data. The present work provides the \lyaf fluctuations catalog.
	\item \textbf{Describe the \lyaf catalog used in the context of early DESI data publications:} Multiple related publications are being published in this context within the Lyman-$\alpha$ working group. The objective of \cite{edrmain} is to obtain the first Lyman-$\alpha$ correlation measurements from DESI early data, testing the current pipeline and data quality, and compare its performance to previous eBOSS DR16 analyses. \cite{edrmocks} provides details on the current status of the different procedures used to build mocks for \lyaf analyses. \cite{edrspurious} characterizes the systematics caused by the DESI instrument on the 3D correlations of the \lyaf. \cite{edrzerrors} studies the impact of redshift errors on the 3D cross-correlation of quasars with the \lyaf. P1D analyses are performed in two different papers: \cite{edr1d} presents the 1 dimensional measurement using Fast Fourier Transform (FFT), while \cite{edr1doptimal} makes use of the Quadratic Maximum Likelihood Estimator (QMLE).
	
	Finally, the current publication provides the Lyman-$\alpha$ fluctuations catalog, as well as its validation.
	The decision to use EDR+M2 data for these analyses was motivated by the need for a larger volume of data than the EDR could offer, which leads to better measurements of the correlation function, constraining power and estimation of systematics.
\end{itemize}
EDR data, including \lyaf fluctuations is available now, including this VAC describing Lyman-$\alpha$ fluctuations. However, EDR+M2 will not be released as a separate piece of data and M2 will be released alongside Year 1 (Y1) data.

\subsection{DESI spectroscopic data}\label{sec:desi_sectroscopic_data}
DESI is the largest ongoing spectroscopic survey, operating on the Mayall 4-meter telescope at Kitt Peak National Observatory \citep{instrument_overview_kp}. DESI consists of 5000 fibers placed in the focal plane and distributed across 10 petals. Each fiber is controlled by a robotic positioner, allowing for quick and automatic positioning. The large number of fibers and the fast cadence provided by the automatic positioners allows DESI to measure up to 5000 spectra every 20 minutes over a $\sim 3^\circ$ field \citep{desi_2016b, focal_plane, corrector}. Fibers carry light from the telescope into a separate room, where it is dispersed by ten spectrographs. Each of them with three cameras each (B, R, Z), covering different wavelength ranges. 

While most of the fibers are assigned to science targets, some are used for calibration. Fibers assigned to the sky are essential for sky subtraction, as they enable the removal of the contribution to spectra caused by light from the background sky, particularly emission lines. Other fibers are assigned to standard stars in order to properly calibrate fluxes. This process is performed by comparing the measured counts in the CCD and the expected fluxes from the standard stars, allowing to estimate fluxes for all the other targets through a process called spectroperfectionism (described in \citealt{spectroperfectionism}). Spectroperfectionism provides spectral fluxes and their variances. For the full description of the DESI spectroscopic reduction pipeline see \cite{Guy-pipeline}.

To help data processing, other pipelines are developed for multiple purposes: optimize the observation strategy \citep{survey_operations}, calculate the exposure time \cite{exposure_time_calc}, assign fibers to targets \cite{fba}, and target selection \citep{ts_pipeline}.

DESI target selection is based on the public Legacy Surveys \citep{bass, ls_overview, ls_dr9}, with preliminary target selection details published for the MWS \citep{mws_preliminary}, BGS \citep{bgs_preliminary}, LRGs \citep{lrg_preliminary}, ELGs \citep{elg_preliminary} and QSOs \citep{qso_preliminary} samples. A series of five papers describe the target selection for these objects (MWS, \citealt{mws_ts}; BGS, \citealt{bgs_ts}; LRG, \citealt{lrg_ts}; ELG, \citealt{elg_ts}; QSO, \citealt{qso_ts}).

Quasars at redshift higher than 2.1 are identified by the DESI pipeline as high redshift quasars suitable for \lyaf studies. These will be re-observed, allowing for a better measurement of their spectra. Re-observations of the same quasar are then coadded and stored in files grouped by healpix pixels (see \citealt{healpix}). The coaddition consists in a simple weighted average of fluxes for each wavelength in the spectra, using as weights their inverse-variance. Coadded files alongside the quasar catalog are used by our pipeline to build the \lyaf catalog.

Each coadded file contains per-spectra information for the three arms (B, Z, R), as well as various metadata, including fiber positions, instrument configuration during observation and atmospheric conditions. The spectroscopic data include fluxes, estimated inverse-variance and a mask identifying invalid pixels.

Each of the three arms of the spectrograph covers a different wavelength region:
\begin{itemize}
	\item B arm: $[3600, 5800]$ \r{A}
	\item R arm: $[5760, 7620]$ \r{A}
	\item Z arm: $[7520, 9824]$ \r{A}
\end{itemize}
with some overlap between each of the arms. The \lyaf is predominantly observed in the blue arm (B arm) of the spectrograph, and only partially in the red arm (R arm). DESI follows a linear wavelength grid with 0.8 \r{A} steps, \textsc{picca} is adapted to work with this resolution.

\begin{figure}
	\centering
	\includegraphics[width=1\columnwidth]{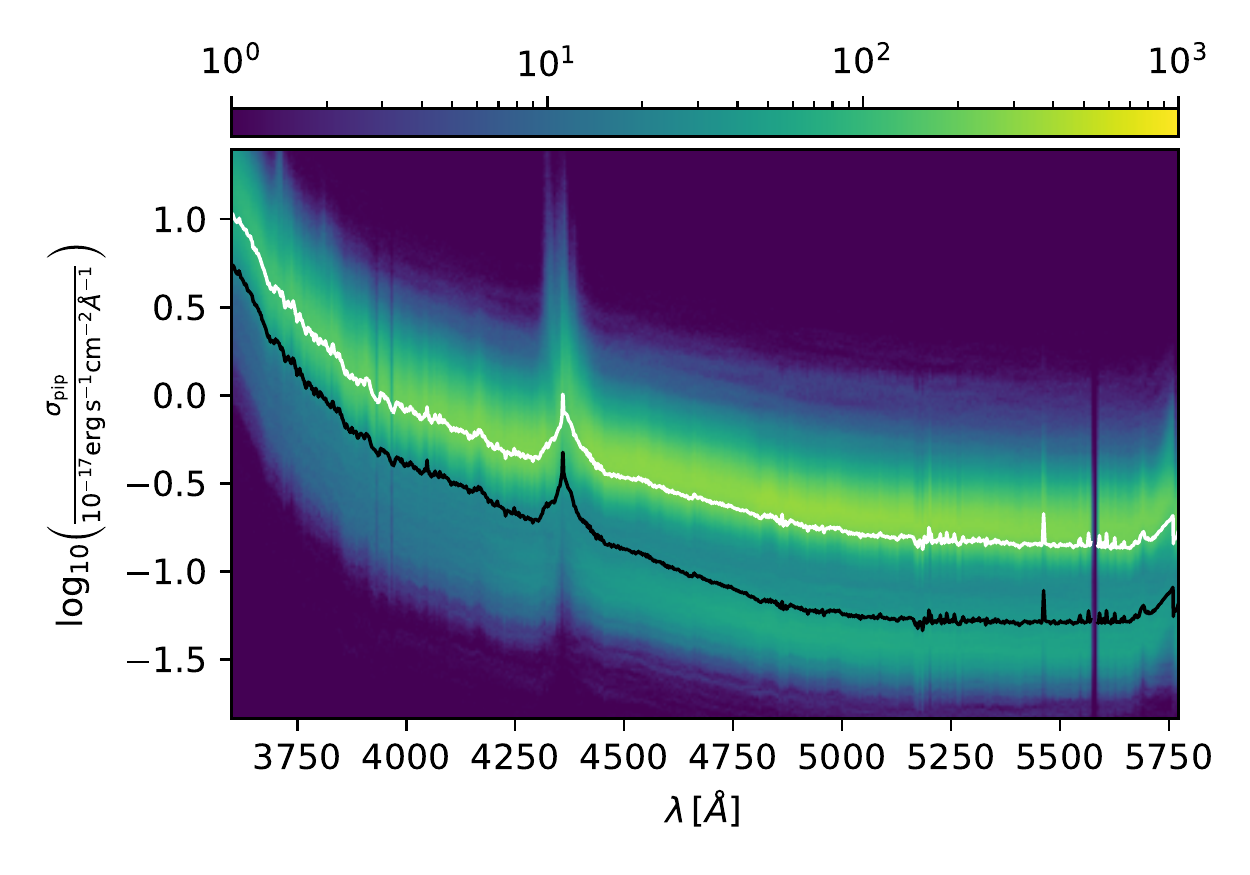}
	\caption{
		Distribution of wavelength and pipeline-reported error on the flux measurement for the larger EDR+M2 sample. This measurement is performed in the \ion{C}{iii} region of the spectra, as defined in Table \ref{table:regions}. The solid lines mark the mean value of the error for a given wavelength, black for the EDR sample and white for the EDR+M2. Apart from the clear distinction in these two lines, a subtle division into two bands at larger wavelengths can also be observed. This is caused by the better signal-to-noise in the EDR sample, thanks to multiple re-observations of the same targets.
		We can also observe how the variance relatively increases in the two ends of the spectrograph and in the area affected by the collimator mirror reflectivity around 4400 \r{A}. 
		}
	\label{fig:desi_noise_and_mask}
\end{figure}

In Figure \ref{fig:desi_noise_and_mask}, we show the distribution of wavelength and pipeline-reported error on the flux measurements for the EDR+M2 dataset, where $\sigma_{\rm pip}$ the variance reported by the DESI pipeline based on photon statistics and readout noise. The image highlights the main differences between EDR and M2 samples. EDR data corresponds to longer observations, with fewer quasars observed. M2 data, on the other hand, includes a large number of objects, but with shorter observations, leading to a larger value for $\sigma_{\rm pip}$. The same panel shows an increase in variance around 4400 \r{A}, caused by the collimator mirror reflectivity \citep{Guy-pipeline}. For reference, the distribution of fluxes is centered around $5 \cdot 10^{-18} {\rm erg} \, {\rm s}^{-1}{\rm cm}^{-2}\AA^{-1}$ in the blue end of the spectra, and $ 5 \cdot 10^{-19} {\rm erg} \, {\rm s}^{-1}{\rm cm}^{-2}\AA^{-1}$ in the red end.

To match the accuracy of our most reliable mock data \citep{edrmocks} and due to the limited amount of \lyaf data available at longer wavelengths, we have limited our analysis to fluctuations fulfilling $z < 3.79$. Given that Lyman-$\alpha$ fluctuations can be related to a specific location in the wavelength grid ($z=\lambda/\lambda_{{\rm Ly}\alpha}-1$), this redshift cut corresponds to a maximum wavelength of 5772 \r{A} entering our analysis.

\subsection{Description of the quasar catalog}
Objects in the spectroscopic data are identified using the template-fitting code Redrock (\citealt{redrock}, and for the special case of QSOs, \citealt{redorck_qso}). Redrock employs a set of templates as representatives of the main object classes observed by DESI: quasars, galaxies and stars. For each observed spectrum, Redrock determines the best-fitting redshift and template by comparing it to the set of templates. This process allows Redrock to accurately identify the objects in the DESI spectroscopic data.

However, Redrock sometimes misidentifies quasars as galaxies. To address this issue, a set of independent quasar-identification approaches, referred to as "afterburners" are also run. These afterburners can identify features missed by Redrock and improve the accuracy of quasar identification. In our case, the most relevant afterburners are QuasarNet \citep{quasarnet,farrquasarnet}, the \ion{Mg}{ii} afterburner, and SQUEzE \citep{squeze}. While only the first two methods were used to build the final catalog, all of them were tested during the SV phase (\cite{vi_qsos}, also see \cite{vi_galaxies} for SV results on galaxies).

For the case of QuasarNET, its main role is to correct for cases where Redrock identifies an object as a low redshift galaxy when it is actually a high redshift quasar. In these cases, QuasarNet is used to re-identify the object as a quasar based on its spectral features using Machine Learning techniques and trained using visually inspected datasets. If QuasarNet identifies an object as a high redshift quasar, Redrock is run again with a high redshift prior. This will likely change the object identification to a high redshift quasar, and if confirmed, the object would be included in the final catalog.

The other relevant afterburner is the \ion{Mg}{ii} afterburner. Its main role is to correct for cases where Redrock misidentifies a quasar as a galaxy. For every object identified as a galaxy by Redrock, the \ion{Mg}{ii} afterburner checks the width of the \ion{Mg}{ii} emission line. If the line is wide enough, the object is re-identified as a quasar and included in the final catalog.

\begin{figure}
	\centering
	\includegraphics[width=1\columnwidth]{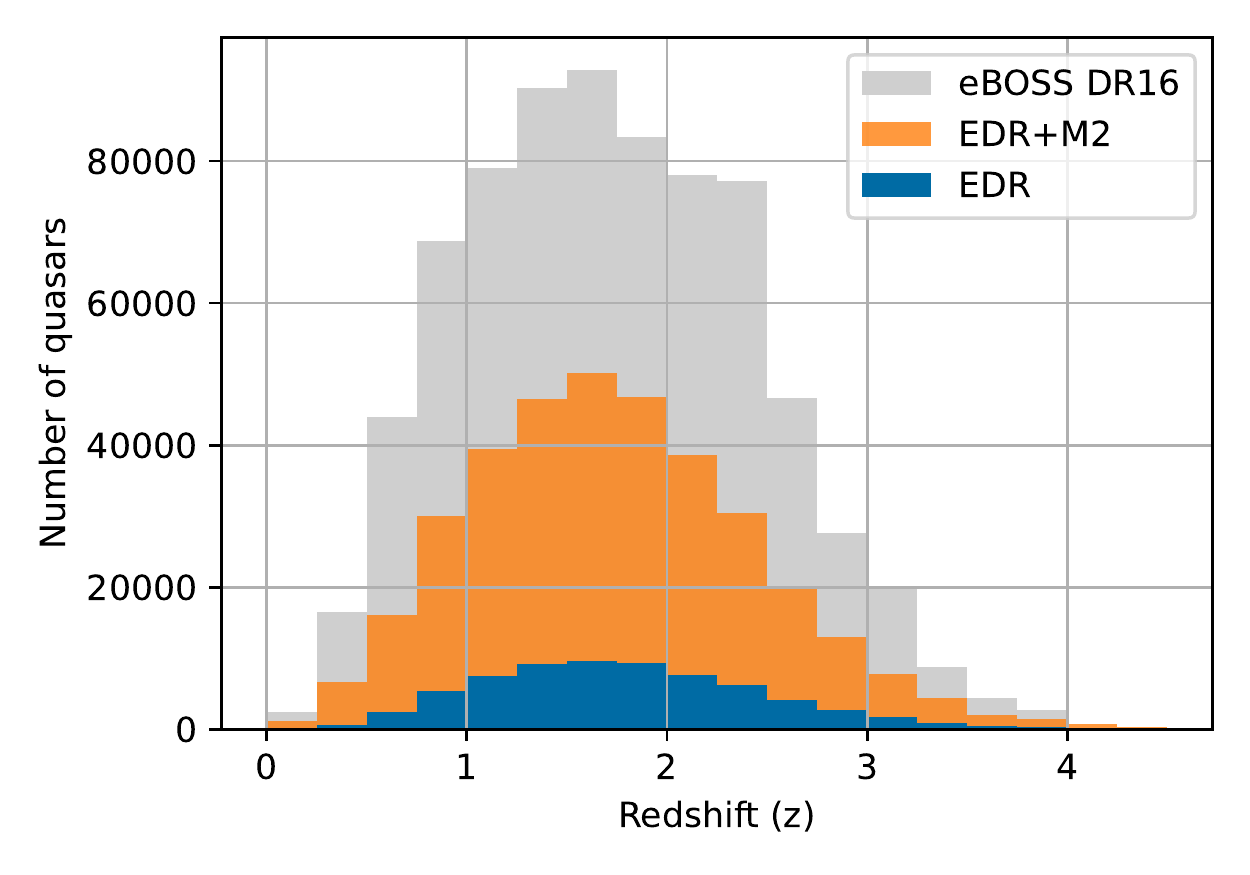}
	\includegraphics[width=1\columnwidth]{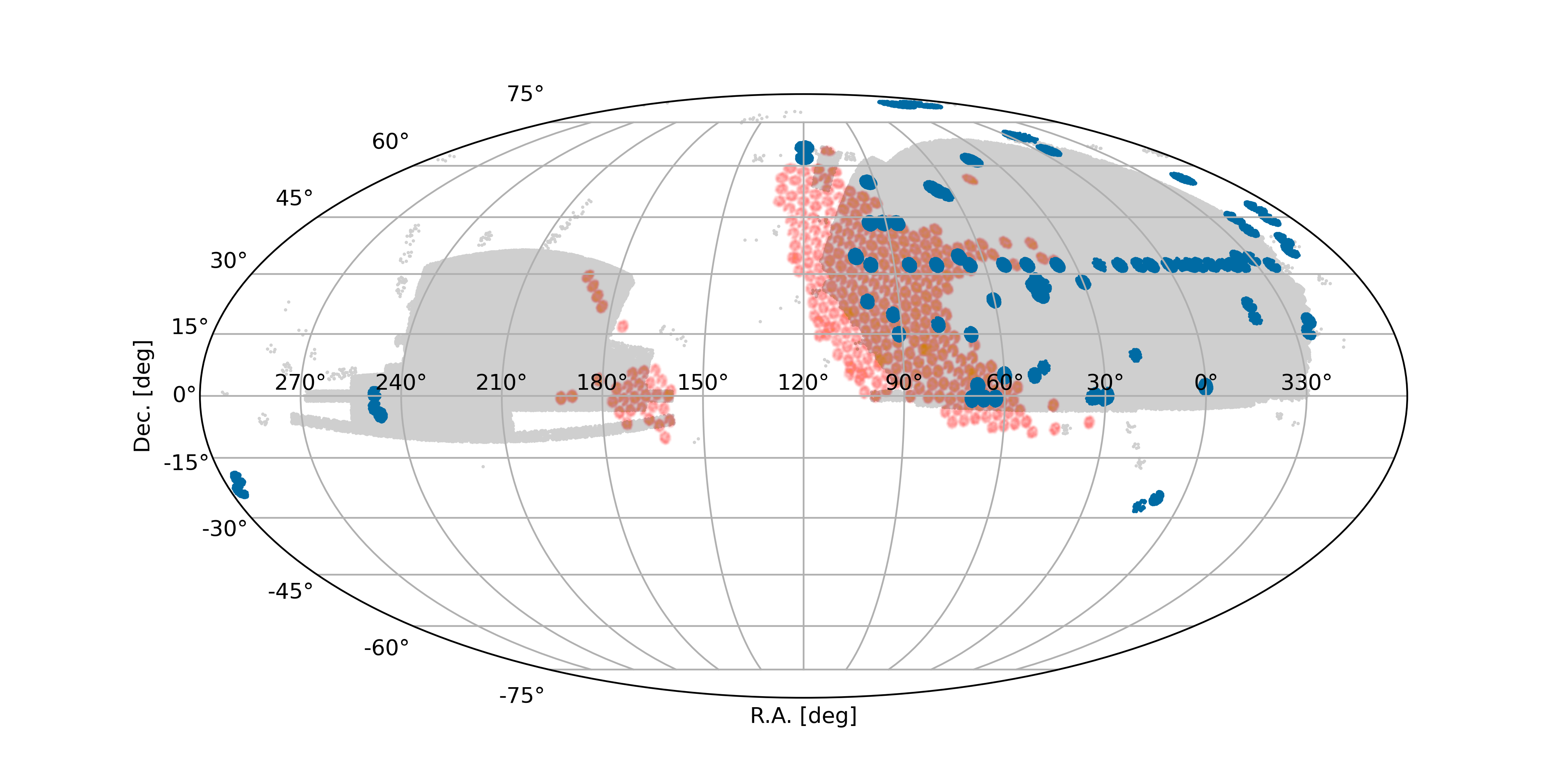}
	\caption{Distribution of objects in the two samples used in this publication, compared with the final eBOSS DR16 sample presented in \protect\cite{dMdB_2020}. Top: Redshift distribution of the three catalogs. Bottom: Sky distribution of the same catalogs. We observe that M2 makes a significant portion of the EDR+M2 sample. When compared to the eBOSS data, the EDR+M2 sample is approximately halfway to matching the number of objects in the former.}
	\label{fig:distribution_of_objects}
\end{figure}

For this release, only objects targeted and confirmed as quasars are used. The resulting catalogs include a total of 68,750 quasars for the EDR sample and 318,691 quasars for the EDR+M2 (see Table \ref{table:forest-sizes}). This value includes all quasars in the input sample, being the actual value of quasars used for each region detailed in Table \ref{table:regions}. The redshift and spatial distributions of the quasars is shown in Figure \ref{fig:distribution_of_objects}, compared to the larger catalog used in eBOSS DR16 analysis \citep{dMdB_2020}. The number of quasars in the EDR+M2 is clearly dominated by the first two months of main survey (M2), containing almost half of the objects in the eBOSS sample.

\subsection{BAL and DLA information}
\label{subsec:balanddlainfo}
Damped Lyman-$\alpha$ Absorption (DLA) systems caused by \ion{H}{i}-rich galaxies affect the \lyaf by generating absorption features that can interfere with the continuum fitting process. DLA information is provided using a DLA finder based on a convolutional neural network and a gaussian process, for a full description of the DLA catalog see \cite{dlajiaqi}. We selected from the full sample the objects where both the convolutional neural network and the gaussian process detected a DLA with a confidence larger than 50\%. %

Apart from DLAs, the \lyaf can also be affected by broad absorption lines believed to be caused by the existence of ionized plasma outflows from the accretion disk. These BAL quasars can be added to the analysis, although they have to be appropriately masked (see Section \ref{sec:masks}). The algorithm for BAL identification was presented in \cite{Guo:2019pal}, and the detailed description of the BAL catalog for DESI data is presented in \cite{edrbal}.

The number of DLAs and BALs for both EDR and EDR+M2 samples is shown in Table \ref{table:forest-sizes}. Given the better signal-to-noise in the EDR sample, the identification of DLA and BAL objects in this sample is higher in this smaller dataset.

\begin{table}
	\centering
	\begin{tabular}{lcc}
		\hline
		& EDR & EDR+M2 \\
		\hline
		Quasars      & 68,750                             & 318,691                             \\
		DLAs      & 5,006                              & 17,375                                 \\
		BALs      & 9,294                              & 28,185                                 \\ \hline
	\end{tabular}
	\caption{
		Number of Lyman-$\alpha$ quasars in the two samples, number of quasars showing BAL features and number of DLAs affecting forests from the Lyman-$\alpha$ quasars. The better signal-to-noise in the EDR sample allows for the detection of a higher number of DLA and BAL features.
	}
	\label{table:forest-sizes}
\end{table}

\section{Spectral reduction}
\label{sec:continuumfittingprocedure}
The continuum fitting procedure estimates the expected flux for each of the quasars in the catalog, this process is essential for computing Lyman-$\alpha$ fluctuations. The flux-transmission field can be defined as:
\begin{equation}
	\delta_q(\lambda) \equiv \frac{f_q(\lambda)}{\overline{F}(\lambda)C_q(\lambda)} - 1 \, ,
	\label{eq:deltafluxdefinition}
\end{equation}
where $\overline{F}$ is the mean transmitted flux at a specific wavelength, $C_q$ the unknown unabsorbed quasar continuum for quasar $q$ and $f_q$ its observed flux. The combination
 $\overline{F}C_q(\lambda)$ is the mean expected flux of the quasars, and is the quantity that we fit for the spectra.

{\it Continuum fitting} is the procedure to compute the flux-transmission field. It can be split in an initial clean-up phase and a second phase where the quasar continuum is actually fitted. The clean-up phase involves two sequential procedures: masking multiple unmodelled features and re-calibrating the spectra to eliminate residuals from the DESI calibration procedure.

In this section we first explain these two procedures, and then provide details on the core of the continuum fitting process.

\subsection{Masks (sky lines, galactic absorption, DLA, BAL)}
\label{sec:masks}
There are multiple features in the spectra that are not caused by Lyman-$\alpha$ fluctuations but are still present in our spectra due to contamination. In order to simplify the cosmological modeling when using Lyman-$\alpha$ fluctuations for correlation measurements, we mask these features and remove them from our analysis.

The three type of masks applied in our analysis are: DESI pipeline; BAL and DLA, applied to absorption features in the forest\footnote{We will refer to each individual line-of-sight as forest, following the convention from previous Lyman-$\alpha$ publications.} region; and galactic absorption and sky emission lines that appear in the spectrograph when observing Lyman-$\alpha$ quasars. We note that the DESI pipeline mask is applied before the individual observations are co-added (see Section~\ref{sec:desi_sectroscopic_data}). The other masks are applied after the coaddition. 

\subsubsection{DESI pipeline masking}
A masking process is applied within the DESI pipeline to identify bad pixels in the spectra. These are typically caused by CCD defects or cosmic rays hitting the spectrograph CCD. This mask is found to only affect about 0.1\% of the DESI pixels used for the Lyman-$\alpha$ analysis. Given the small fraction of pixels discarded, we decided to remove these pixels from the analysis.

\subsubsection{DLA and BAL masking}
\label{subsubsec:balanddlamask}
DLA absorption affects our spectra by imprinting itself in the spectra of quasars, resulting in zero flux around the true redshift of the DLA and damping wings further away. A secondary effect of this is the reduction in the mean flux of the affected spectra, affecting the overall mean absorption. This biases our estimate of $\overline{F}C_q$, potentially affecting all the quasars in the sample (see Section \ref{subsec:continuumfitting}).

Although it is possible to include the absorption features from DLAs into our models for the correlation functions, for simplicity, and following the prescription in previous analyses \citep{dMdB_2020}, we mask the regions of the \lyaf that are affected by identified DLAs when the DLA reduces the transmission by more than 20\%. We then correct the absorption in the wings using a Voigt profile as suggested by \cite{Noterdaeme}, being able to include the pixels affected by these wings without affecting the mean absorption.

\begin{figure}
	\centering
	\includegraphics[width=1\columnwidth]{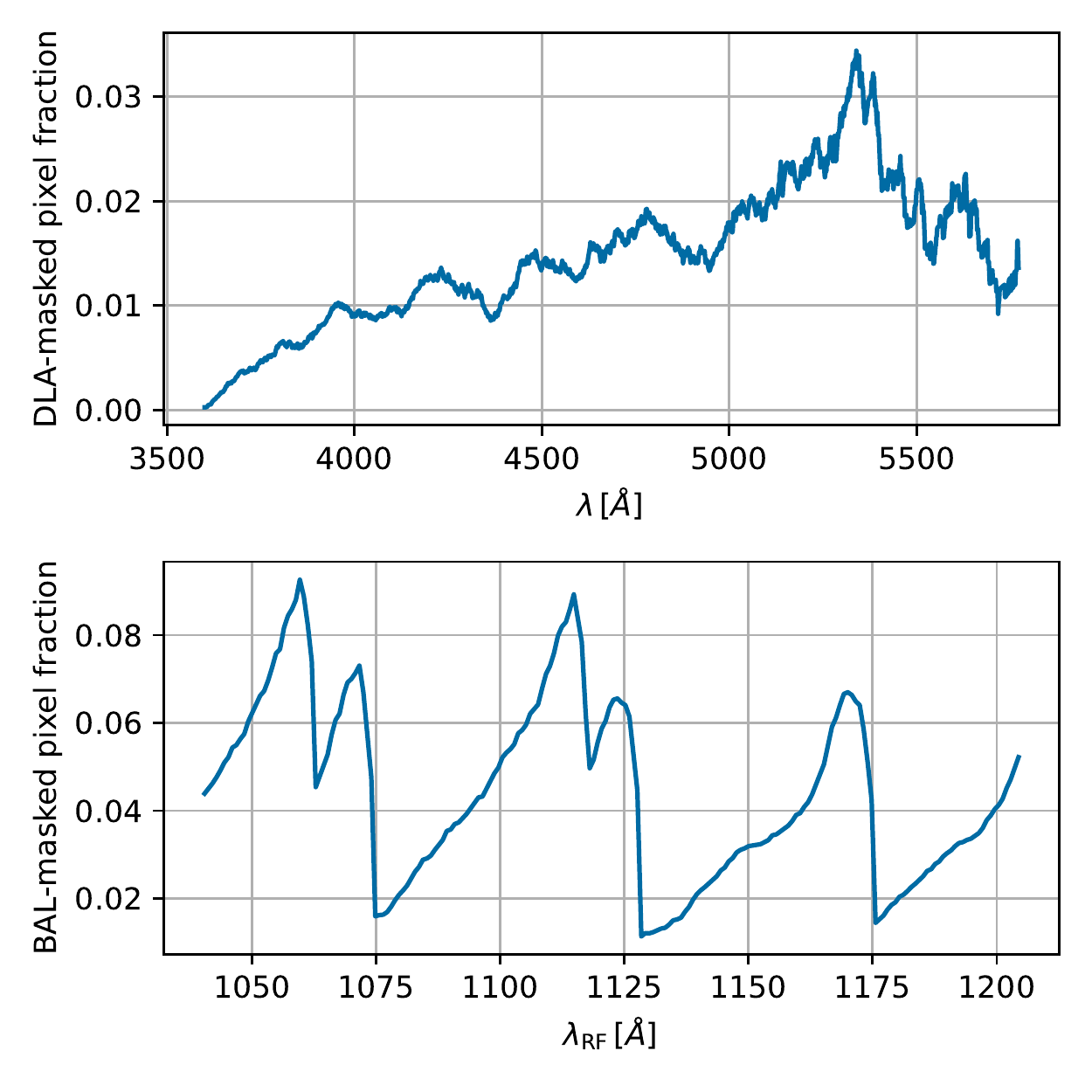}
	\caption{
		Fraction of pixels masked due to DLAs (top) and BAL (bottom) features. As expected, the number of detected DLAs increase with redshift (and therefore with $\lambda$), yielding a fraction of masked pixels always below the 5\%. For the BAL case, we show the masked fraction as a function of $\lambdarf$, which allow us to observe the strong wavelength dependence of the masking, associated with emission lines for different elements. This Figure used the full EDR+M2 dataset.
	}
	\label{fig:dla_bal_pixel_fraction}
\end{figure}

The top panel of Figure \ref{fig:dla_bal_pixel_fraction}, shows the fraction of pixels that have been masked due to DLAs as a function of observed wavelength. Although there is an increase in the number of detected DLAs with redshift, the number of masked pixels is always smaller than 5\%. 

Regarding BALs, in previous analyses quasars exhibiting BAL features were directly excluded from the analysis, although they were later used as tracers for cross-correlations between quasars and the \lyaf. However, \cite{BALsEnnesser} showed that quasars with BAL features could be safely included in the analysis if all expected locations of these absorption features are masked. Here, we follow their proposed approach.

In the bottom panel of Figure \ref{fig:dla_bal_pixel_fraction}, we observe the fraction of pixels masked due to BALs as a function of rest-frame wavelength. In this case, we can see a pattern of absorption that matches the expected absorption features described in \cite{BALsEnnesser}. Since not all the BAL quasars suffer from the same absorption profile, we expect the BAL-masked pixel fraction to be maximal around the central absorption and decrease as we go away from it. It is worth noting that for the quasars used in the Lyman-$\alpha$ region continuum fitting, the percentage of BAL quasars is 16\% for the EDR+M2 sample and 23\% for the EDR sample.	

\subsubsection{Galactic absorption and sky emission masking}
There are certain absorption and emission features in our spectra that differ from Lyman-$\alpha$ fluctuations or other absorbers in the IGM. These absorption and emission features can be easily identified because they affect specific wavelengths, resulting in sharp features when the spectra of multiple quasars is combined. We can remove them by simply masking out the corresponding wavelengths from our analysis.

\begin{figure}
	\centering
	\includegraphics[width=1\columnwidth]{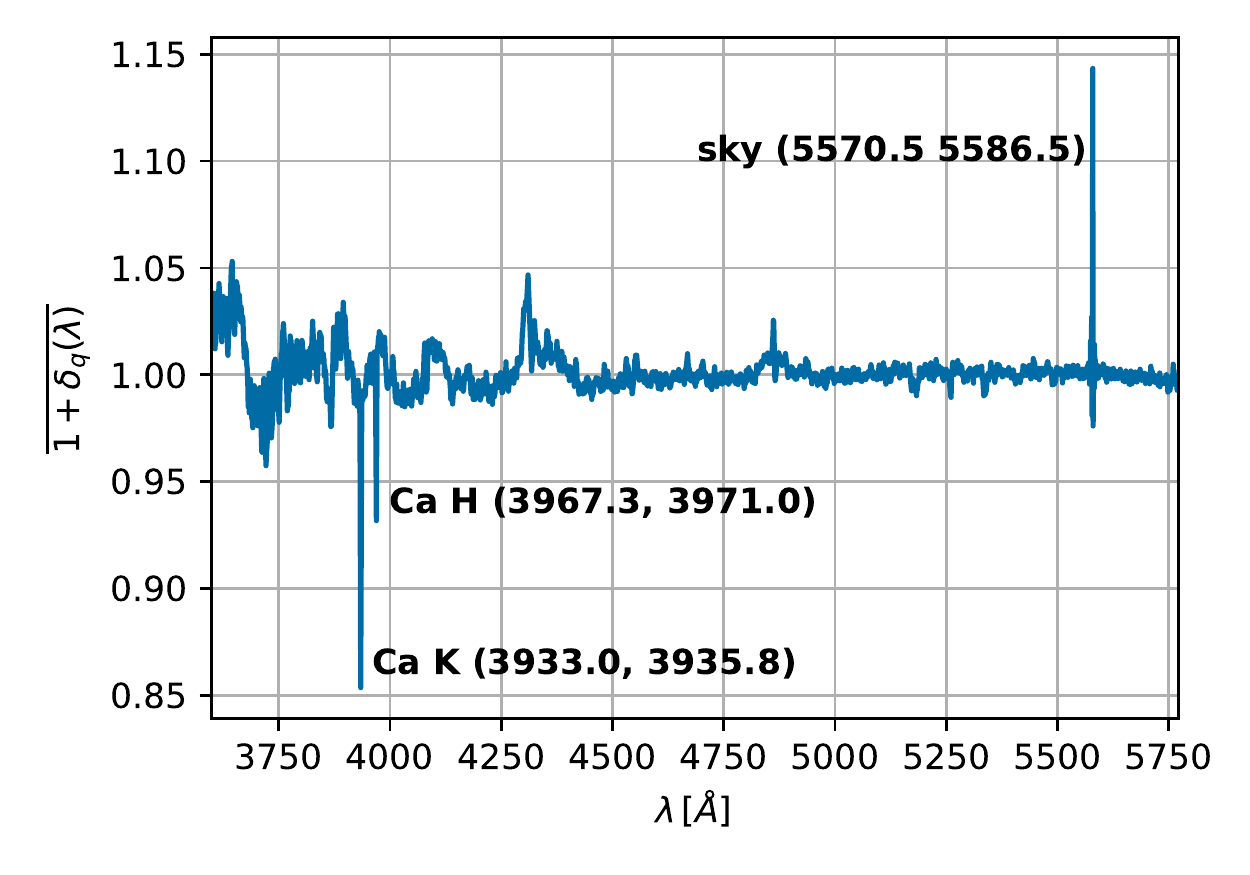}
	\caption{
		Weighted average of the flux-transmission field measured in the \ion{C}{iii} region. This measurement has been performed without masking sharp features, and they can be clearly observed at different positions in the spectrograph. The three masked regions are specified in the plot, showing the wavelength range affected by the mask. Smooth features in this measurement can also be observed. Their impact can be corrected through the flux re-calibration (see Section \ref{sec:calibration}). For this Figure we used the full EDR+M2 dataset.
	}
	\label{fig:masks_due_to_absorption_and_emission}
\end{figure}

The two most significant features in this regard are galactic absorption and sky emission. Galactic absorption is caused by material in the Interstellar Medium (ISM) absorbing at specific wavelengths. The ISM absorption for some of these lines cannot be easily separated from the intrinsic absorption in the atmospheres of the stars used for flux calibration, and thus they are not properly accounted for. Here, we are affected by the Ca K and H transitions, and we mask the corresponding wavelengths (see Figure~\ref{fig:masks_due_to_absorption_and_emission}).

Sky emission comprises emission lines generated by atmospheric effects and are mostly corrected in the modelling of sky lines. However, inaccuracies in this modelling result in spurious features in our spectra. Here, we also mask the affected wavelengths (see figure~\ref{fig:masks_due_to_absorption_and_emission}).

In Figure \ref{fig:masks_due_to_absorption_and_emission}, we present the measurement of the estimated $\overline{1 + \delta_q(\lambda)}$ in the \ion{C}{iii} region (see Table \ref{table:regions}). Two kind of features can be observed in this plot: first, we observe the sharp features corresponding to the mentioned galactic absorption and sky emission; then we observe smooth features caused by inaccuracies in the DESI calibration process. To correct for the former, we mask the pixels in the wavelength intervals shown in the plot. The later can be corrected through re-calibration (see Section \ref{sec:calibration}). 

Here we use the \ion{C}{iii} region to build our masks as it lacks the Lyman-$\alpha$ absorption features. This makes it easier to estimate absorption and emission effects. %

\subsection{Re-calibration}
\label{sec:calibration}
DESI calibrates fluxes using standard stars \citep{Guy-pipeline}. However, inaccuracies in the modelling of these calibration stars introduce features in the measured spectra when fluxes are predicted for other objects (such as quasars). These features can be observed by examining the mean $\delta_q(\lambda)$ in any region of the spectra, particularly in the regions without Lyman-$\alpha$ absorption, where the variance of the measured flux is expected to be lower. This is possible because the features caused by calibration defects are function of observed wavelength, while the continuum estimate is a function of rest-frame wavelength. Looking again at Figure \ref{fig:masks_due_to_absorption_and_emission}, we can observe smooth features in the stack of fluctuations ($\overline{1+\delta_q(\lambda)}$) alongside the masked sharp features.

These features are correlated between different forests, and could therefore bias the measurement of the correlation function. To avoid this, on top of the DESI pipeline calibration, we re-calibrate our spectra. As stated above, these features affect the same region of observed wavelength regardless of the rest-frame position where they act. Therefore, one could in principle use the entire quasar spectra for calibration purposes. However, the larger spectral diversity near quasar emission lines can unnecessarily complicate this. In consequence, we search for a featureless region redwards of the Lyman-$\alpha$ emission line. We will refer to all the candidate regions as re-calibration regions.

In the absence of Lyman-$\alpha$ fluctuations, flux fluctuations are ideally only caused by noise. In any of the re-calibration regions, the measurement of $\overline{1 + \delta_q(\lambda)}$ is expected to be consistent with 1; and its measurement can be considered a null test that is not fulfilled in general given the issues described above. 

\begin{figure}
	\centering
	\includegraphics[width=1\columnwidth]{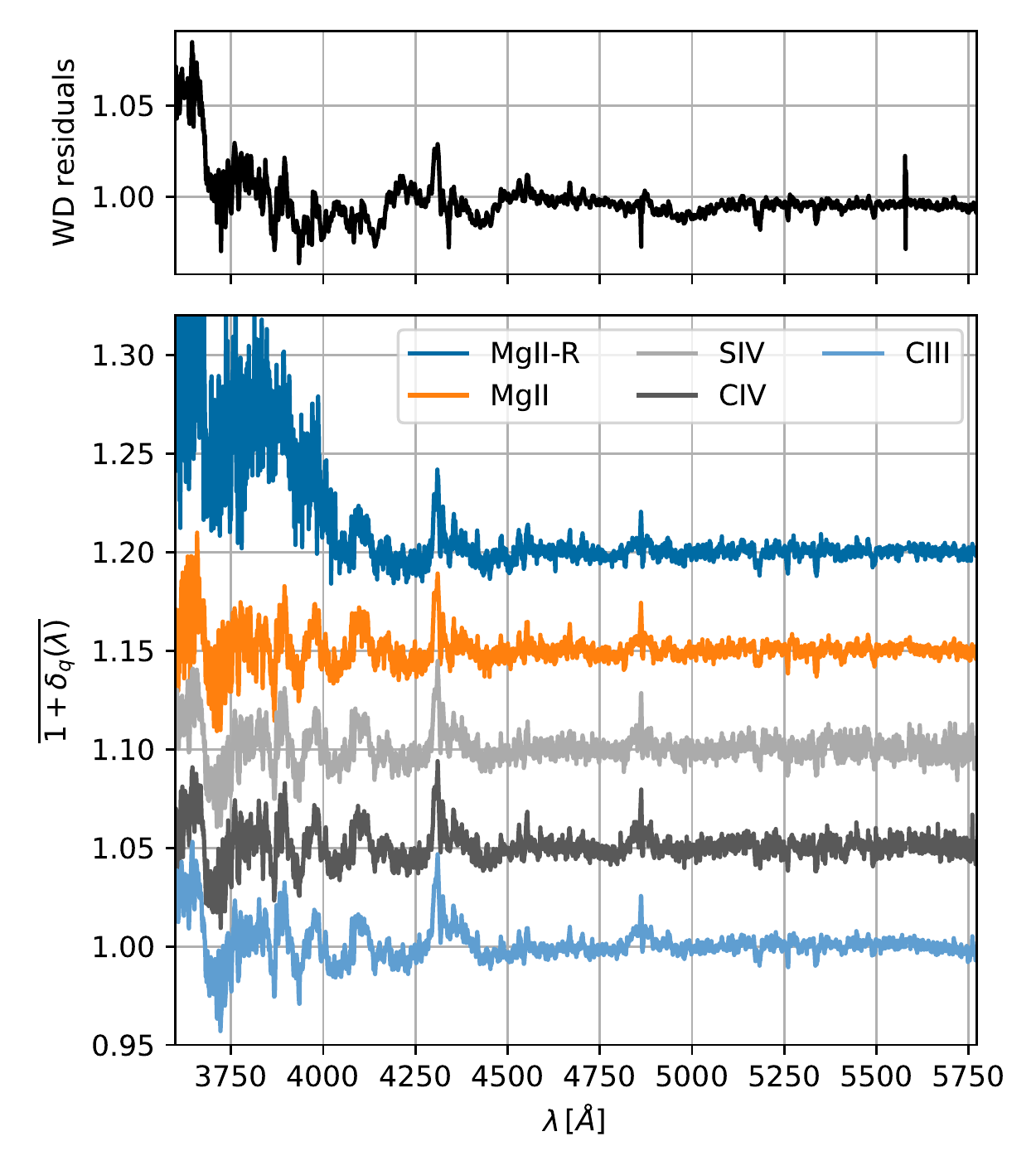}
	\caption{
		Bottom: Weighted average of the flux-transmission field measured at different regions of the spectra. Its value has been shifted to better distinguish the different regions. As opposed to the results shown in Figure \ref{fig:masks_due_to_absorption_and_emission}, sharp features are not present here because these samples have already been masked. The smooth features in the spectra are similar for all the measured regions, being the \ion{Mg}{ii}-R an outlier in this tendency, likely to be caused due to the reduced number of pixels available for this region at low wavelengths (see Figure \ref{fig:n_pixels}). Results are computed from the full EDR+M2 sample. Top: Average residuals to White Dwarf (WD) spectra in the blue arm of the DESI spectra, as seen in \protect\cite{Guy-pipeline}. A similar trend can also be observed here between these residuals and our measured average of the flux-transmission field, especially at smaller wavelengths, further justifying our re-calibration process. 
	}
	\label{fig:calib_regions_comparison}
\end{figure}

In the bottom panel of Figure \ref{fig:calib_regions_comparison}, we show the measurement of $\overline{1 + \delta_q(\lambda)}$ for multiple different candidate re-calibration regions. The fact that all of them show similar features, in spite of being at different regions of the spectra, suggests that all fluxes can be corrected consistently. Furthermore, we see comparable features in the White Dwarf average residuals in the top panel of the same Figure. This similarity can be seen in some of the regions of the spectrum, especially for $\lambda \in [3600, 4200]$ \r{A}, and it further justifies the re-calibration process.

\begin{figure}
	\centering
	\includegraphics[width=1\columnwidth]{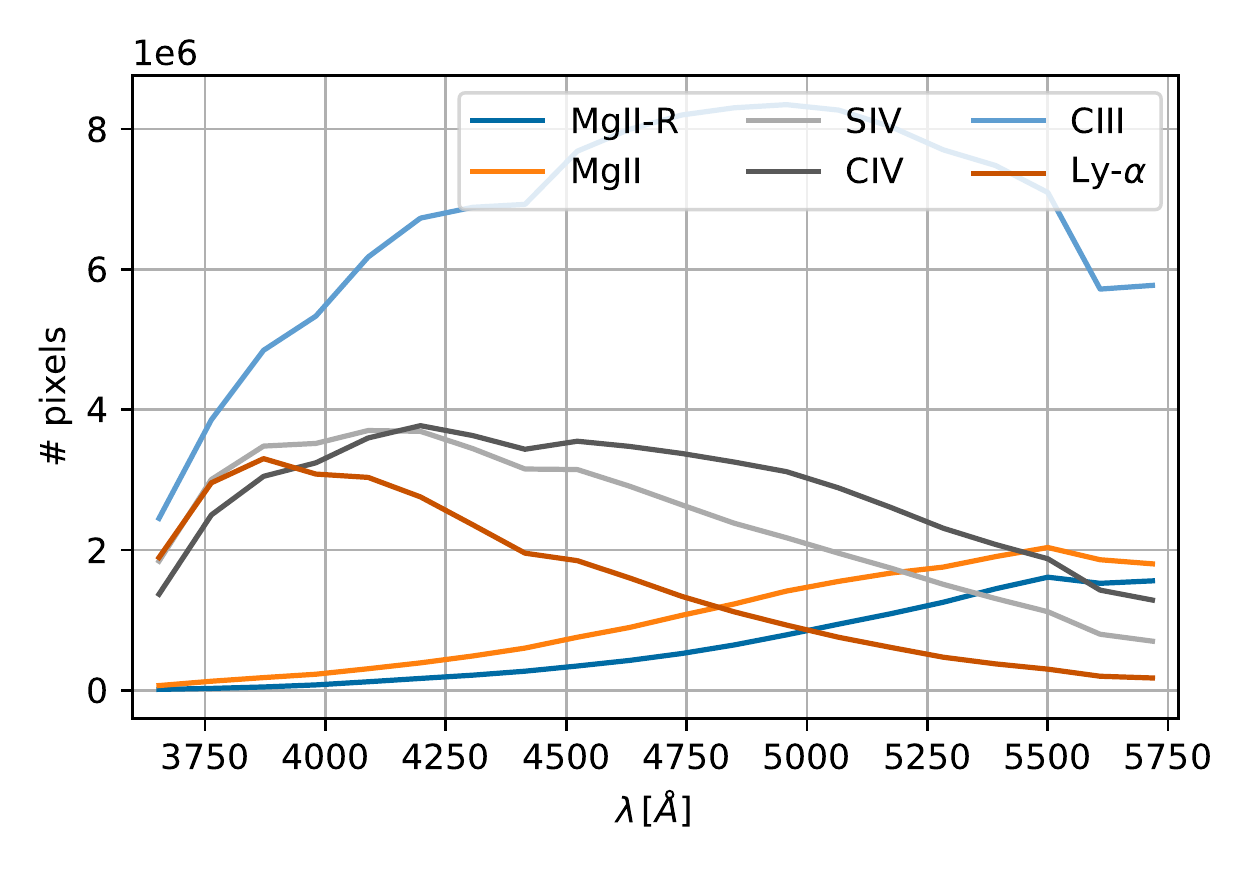}\
	\caption{
		Number of pixels available for the different regions measured in bins of $\Delta \lambda = 55.58$ \r{A}. Given the location of the different regions at different positions in the quasar spectra, the number of available pixels is different for each of them. This is because our spectral coverage lies in the range $(3600, 5772)$ \r{A}. Thanks to the larger number of pixels available for the \ion{C}{iii} region, it was selected as the region to be used for the re-calibration process. This Figure used the full EDR+M2 sample.
	}
	\label{fig:n_pixels}
\end{figure}

In previous analyses, a region to the right of the \ion{Mg}{ii} emission line (\ion{Mg}{ii}-R) was selected to re-calibrate fluxes, partly because it is located further to the right of the spectra, reducing potential contamination from other absorption lines. However, in this work we selected the \ion{C}{iii} region ($\lambda \in [1600, 1850]$ \r{A}) due to the larger number of pixels available and given the similar behavior compared to the other regions. Figure \ref{fig:n_pixels} shows the number of pixels at each wavelength bin of size $\Delta \lambda = 55.58$ \r{A} for the different regions, and the number of forests available for each region can be found in Table \ref{table:regions}. In both cases, we see that \ion{C}{iii} has the largest number of pixels available for the analysis.

\begin{table*}
	\begin{tabular}{lcccc}
		\hline
		Region & $\lambda_{\rm RF, min}$ & $\lambda_{\rm RF, max}$ & \# forests EDR & \# forests EDR+M2\\
		& \r{A} & \r{A} & & \\ \hline
		\ion{C}{iii}                                  & 1600                                                          & 1850                                                          & 49,810                                   & 233,310                                       \\
		\ion{C}{iv}                                   & 1410                                                          & 1520                                                          & 41,445                                   & 189,984                                       \\
		\ion{Mg}{ii}-R                                  & 2900                                                          & 3120                                                          & 7,290                                   & 33,936                                        \\
		\ion{Mg}{ii}                              & 2600                                                          & 2760                                                          & 13,373                                   & 62,628                                        \\
		\ion{S}{iv}                                   & 1260                                                          & 1375                                                          & 34,044                                   & 152,979                                       \\
		Lyman-$\alpha$                        & 1040                                                          & 1205                                                          & 20,281                                        & 88,511                                        \\ \hline
		\end{tabular}
	\caption{Statistics for the regions considered during the analysis, the region span is defined in the quasar rest-frame wavelength ($\lambdarf$). The number of forests corresponds to the number of quasars whose spectra can be observed in the spectrograph. A minor number of forests are rejected due to low signal-to-noise ratio (SNR) or due to them being too short.}
	\label{table:regions}
\end{table*}

The choice of a re-calibration region at a larger wavelength than the \lyaf also leads to an increase in the number of quasars that can be used for this process. The \lyaf analysis includes quasars with redshifts in the range $z \in (2.1, 3.7)$, while for the \ion{C}{iii} region, quasars in the range $z \in (0.9, 2.6)$ are included. By looking at the quasar distribution (Figure \ref{fig:distribution_of_objects}), we see that the number of objects in the second range will be larger, and hence a larger number of pixels available in the re-calibration region.

We use a similar procedure as \cite{dMdB_2020} for this re-calibration, correcting all the fluxes by using the mean flux in the re-calibration region:
\begin{equation}
	f_{\rm new}(\lambda_i) = f_{\rm orig}(\lambda_i) / \overline{f_{\rm calib, orig}}(\lambda_i)~.
	\label{eq:calibration_flux}
\end{equation}
The error reported by the pipeline for these fluxes also needs to be corrected through:
\begin{equation}
	{\sigma_{\rm pip, new}}(\lambda_i) = \sigma_{\rm pip, orig}(\lambda_i) / \overline{f_{\rm calib, orig}}.
	\label{eq:calibration_ivar}
\end{equation}
After this correction, smooth features in the spectra will be alleviated, which prevents biased results. In previous analyses, a second correction was applied to correct the estimation of the flux variance provided by the pipeline (see Appendix \ref{ap:dmdb20procedure}). However, the better performance of DESI in this aspect allows us to skip this step. Evidence in this direction is the better behavior of the estimated correction to the pipeline error (see Figure \ref{fig:delta_attributes}).

\subsection{Continuum fitting}
\label{subsec:continuumfitting}
The core of the continuum fitting process consists of obtaining the expected flux $\overline{F}C_q$ for each quasar in the sample. This allows for the determination of the flux-transmission field (Eq. \ref{eq:deltafluxdefinition}). The process is performed iteratively, with several quantities fitted simultaneously, and performed on each of the quasar regions independently: in the case of a re-calibrated Lyman-$\alpha$ analysis, it is firstly performed in the re-calibration region (\ion{C}{iii} in this case), and afterwards in the Lyman-$\alpha$ region.

In order to simplify the process, the expected flux $\overline{F}(\lambda)C_q(\lambda)$ is assumed to be a universal function of rest-frame wavelength, $\bar{C}(\lambdarf)$, corrected by a first degree polynomial in log $\lambda$.:
\begin{equation}
	\overline{F}(\lambda) C_q(\lambda) = \overline{C} (\lambdarf) \left(a_q + b_q \frac{\Lambda - \Lambda_{\rm min}}{\Lambda_{\rm max}-\Lambda_{\rm min}}\right),
	\label{eq:meanexpectedflux}
\end{equation}
where $\Lambda \equiv \log \lambda$ and $\Lambda_{\rm min, max}$ identify its minimum and maximum values inside the region to be fitted. That is, the spectra of quasars in our sample are assumed to have the same underlying shape or continuum for all objects, allowing for variations in the amplitude and tilt. In the analysis, the transmitted flux $\overline{F}$ and the quasar continuum $C_q$ cannot be fitted independently, although they are not needed separately in our analysis. For this reason, we choose to directly estimate the expected flux of quasars. 

The parameters $(a_q, b_q)$ are fitted by maximizing the likelihood function
\begin{equation}
	2 \ln L = - \sum_i \frac{ \left[ f_i - \overline{F}C_q(\lambda_i, a_q, b_q) \right]^2}{\sigma^2_q (\lambda_i)} - \sum_i \ln \left[ \sigma^2_q (\lambda_i) \right],
	\label{eq:likelihood}
\end{equation}
where $\sigma^2_q(\lambda)$ is the variance of the flux $f_i$. This value has to be estimated, and since it depends on $(a_q, b_q)$, we include this dependence in the likelihood function.

The full variance of the flux, $\sigma^2_q(\lambda)$, includes not only the obvious contribution from the noise estimated by the DESI pipeline but also the intrinsic variance of the \lyaf\footnote{This quantity is expected to be very small outside the \lyaf region.}. We account for the intrinsic variance in the following way:
\begin{equation}
	\frac{\sigma^2_q(\lambda)}{\left( \overline{F}C_q(\lambda) \right)^2} = \eta(\lambda) \tilde{\sigma}^2_{\rm pip, q} (\lambda) + \varlss (\lambda).
	\label{eq:variance_scheme}
\end{equation}
where $\varlss$ is the mentioned intrinsic variance of the \lyaf, and $\tilde{\sigma}_{\rm pip, q} = \sigma_{\rm pip, q}(\lambda)/\overline{F}C_q(\lambda)$ where $\sigma_{\rm pip,q}$ is flux variance as estimated by the DESI pipeline. In this expression we also include a correction $\eta(\lambda)$ to account for inaccuracies in the pipeline noise estimation. We discard here an extra term in this expression accounting for quasar variability at high SNR, which was previously used in \cite{dMdB_2020} (see \ref{app:variance_estimator} for details). It is worth noting that $\sigma^2_q$ here is the variance of the flux and therefore has flux units, whether all quantities at the right hand side are dimensionless.

\begin{figure}
	\centering
	\includegraphics[width=1\columnwidth]{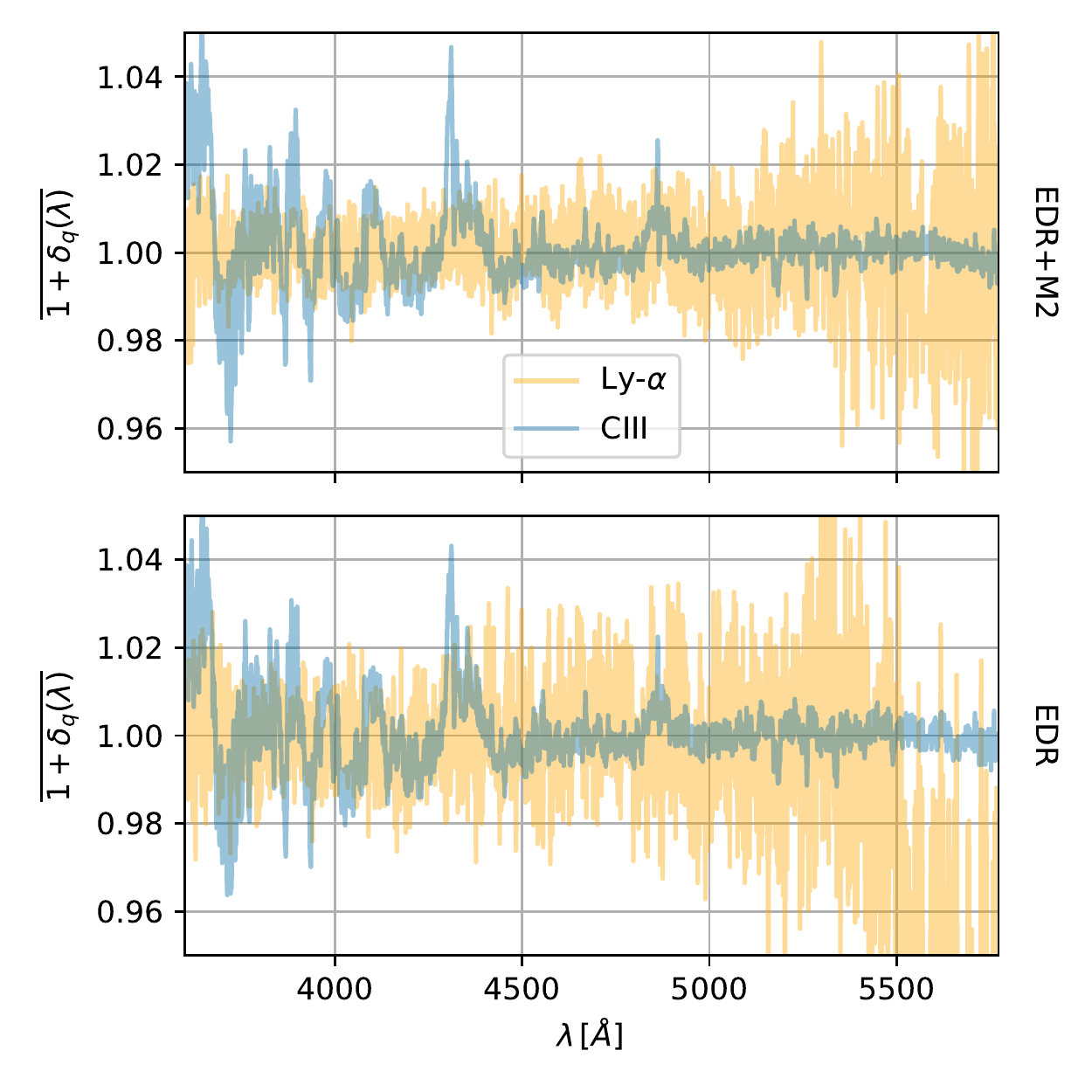}
	\caption{
		Measurement of the weighted mean of the flux-transmission field for both the Lyman-$\alpha$ and \ion{C}{iii} regions. The Lyman-$\alpha$ region shows an expected higher variance at all wavelengths compared to the \ion{C}{iii} region, although it lacks smooth features thanks to the re-calibration process. Similarly, the EDR sample shows a higher variance than EDR+M2. In both cases, this is caused by the decrease in number of pixels available. Given the low number of pixels available at larger wavelengths for the EDR sample for the Lyman-$\alpha$ region, it departs from the expected unity behavior.
	}
	\label{fig:delta_stack}
\end{figure}

In the iterative continuum fitting process, we estimate the quantities $\eta$, $\varlss$, $a_q$, $b_q$, $\overline{C}(\lambdarf)$. The sequential steps of the process are:
\begin{enumerate}
	\item \textbf{Assume wavelength-independent values for unknown quantities:} A flat assumption is assumed for all the quantities to be fitted and measured. In practice, this means setting $\overline{C}(\lambdarf)=1$, $\eta(\lambda)=1$ and $\varlss(\lambda) = 0.1$. \label{step1}
	\item \textbf{Fit the per-quasar parameters $(a_q, b_q)$:} Using the current values of $\overline{C}(\lambdarf)$, $\eta(\lambda)$ and $\varlss(\lambda)$, the pair of values $(a_q, b_q)$ is estimated for each individual forest through the minimization of the likelihood defined in Eq. \ref{eq:likelihood}. \label{step2}
	\item \textbf{Estimate the flux-transmission field $\delta_q(\lambda)$:} Using the current value $\overline{C}(\lambdarf)$ and the best-fit parameters for $(a_q, b_q)$, we compute the expected flux for each quasar following Eq. \ref{eq:meanexpectedflux}. This allows for the computation of the fluctuations around the expected flux ($\delta_q(\lambda)$) as defined in Eq. \ref{eq:deltafluxdefinition}.
	\item \textbf{Fit the variance functions:} The functions $\eta(\lambda)$ and $\varlss (\lambda)$, defined in Eq. \ref{eq:variance_scheme}, are now fitted using the estimated $\delta_q(\lambda)$ from the previous step. In order to do so, different values of the flux-transmission field are grouped by wavelength and $\tilde{\sigma}_{\rm pip}$. We take 20 bins for the wavelength split and 100 for the split on $\tilde{\sigma}_{\rm pip}$, generating a grid of 2000 points. We compute the variance $\sigma^2(\lambda, \tilde{\sigma}_{\rm pip}) = \overline{\delta_q^2(\lambda, \tilde{\sigma}_{\rm pip})}$ for each point in the grid\footnote{Here we assume $\overline{\delta_q(\lambda, \tilde{\sigma}_{\rm pip})}=0$ as an approximation. Fluctuations around this assumption are small (see Figure \ref{fig:delta_stack})}. The functions $\eta(\lambda)$ and $\varlss$ are fitted for each of the 20 wavelengths independently using the following likelihood function:
	\begin{equation} 
		2 \ln L = - \sum_{\tilde{\sigma}_{\rm pip}} \frac{\delta_q^2(\lambda, \tilde{\sigma}_{\rm pip}) - \eta(\lambda)\tilde{\sigma}^2_{\rm pip} - \varlss(\lambda)}{\delta_q^4(\lambda, \tilde{\sigma}_{\rm pip})} ~,
	\end{equation}
where the sum in $\tilde{\sigma}_{\rm pip}$ include all the valid bins in $\tilde{\sigma}_{\rm pip}$ (at most 100). The points in the grid computed using less than 100 pixels are considered unreliable and discarded from the fits. We note that in those cases where the fit does not converge (for example, if there are too few quasars), then the default values from step \ref{step2} are kept.
	\item \textbf{Recompute the mean expected flux:} At this stage we update the value of $\overline{C}(\lambdarf)$. This is performed by computing the weighted average of all quasars expected flux sharing the same $\lambdarf$ value. Here, we use the optimal weights as defined in  Eq. \ref{eq:weight_scheme_alt} (see section~\ref{subsec:optima_weights}).
	\item \textbf{Compute and save relevant statistics:} Relevant statistics are computed at each iteration and stored in \texttt{delta\_attributes\_iteration\{i\}.fits.gz} files (where $i$ is the iteration number). The saved statistics include the stack of fluctuations ($\overline{1+\delta_q(\lambda)}$), the fitted variance functions ($\eta$, $\varlss$), the mean continuum ($\overline{C}(\lambdarf)$) and fit metadata including the tilt and slope values ($a_q$, $b_q$), the number of pixels used for the fit and the $\chi^{2}$ of the fit.
	\item \textbf{Continue next iteration starting in step \ref{step2}:} The next iteration starts using the updated values of $\overline{C}(\lambdarf)$, $\eta$ and $\varlss$.
\end{enumerate}
This process is performed 5 times, resulting in stable estimates of all the defined quantities.

\begin{figure}
	\centering
	\includegraphics[width=1\columnwidth]{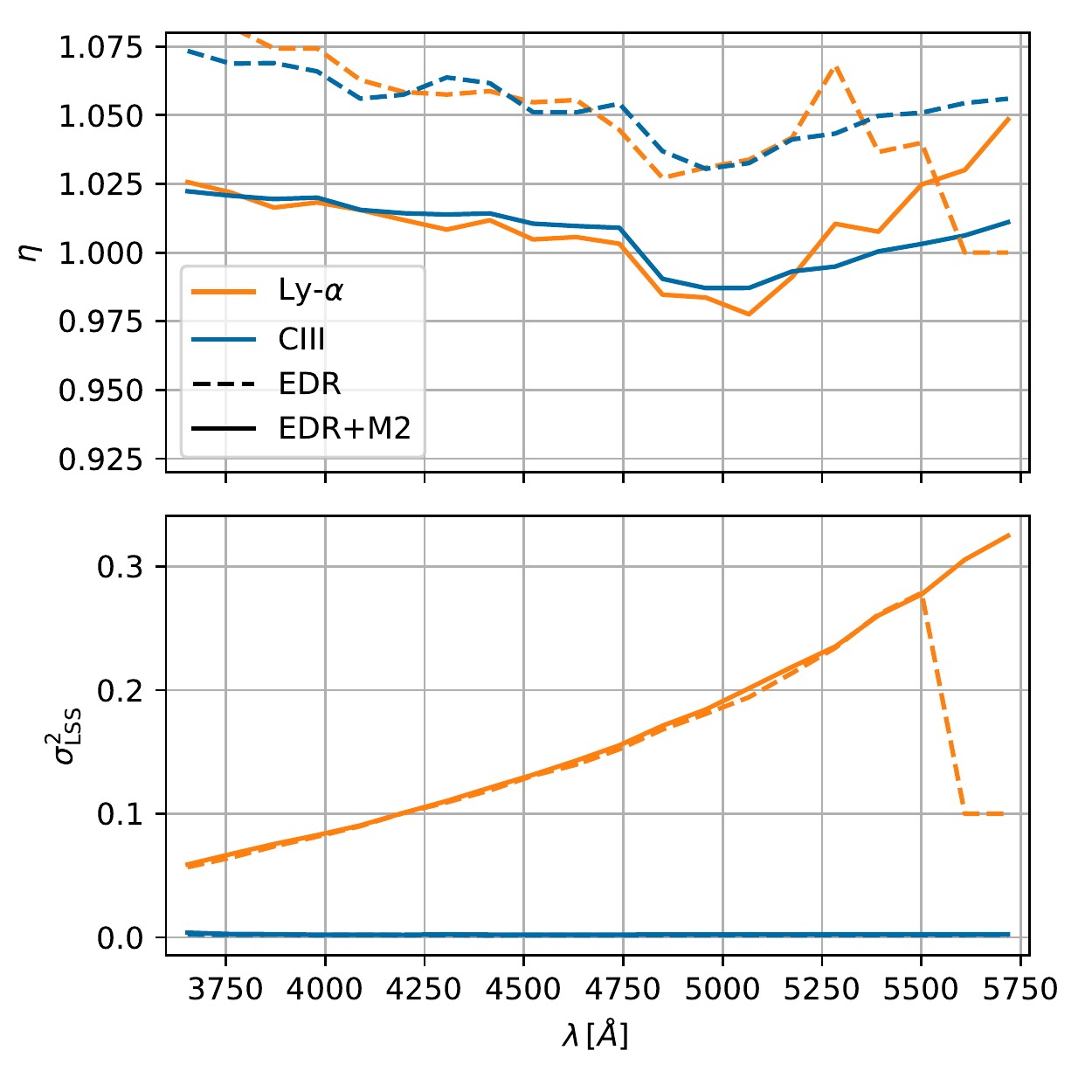}
	\caption{
		Wavelength evolution of the fitted parameters $\eta$ and $\varlss$, measured in both the \ion{C}{iii} and Lyman-$\alpha$ regions for the EDR (dashed) and EDR+M2 (solid) samples. Top: The pipeline error correction $\eta$ is found to be larger for the EDR sample, caused by the worse estimation of the pipeline-reported variance for the first part of the SV program. Bottom: $\varlss$, in this case it is consistent between the two samples. As expected the \ion{C}{iii} region shows a value close to 0 for all wavelengths, while for the Lyman-$\alpha$ region follows the expected increase in its intrinsic variance with redshift. In both $\eta$ and $\varlss$, measurements for the EDR sample, the fitted parameters could not be obtained for the larger wavelength value due to the reduced number of pixels available, falling to the default values $\eta=0.1$ and $\varlss = 0.1$.
	}
	\label{fig:delta_attributes}
\end{figure}

In Figure \ref{fig:delta_attributes}, we show the final estimates for the two fitted function $\eta$ and $\varlss$ for the two regions considered in this analysis (Lyman-$\alpha$ and \ion{C}{iii}) and for the two different samples (EDR and EDR+M2). The correction to the pipeline estimated variance, $\eta$, shows a $\sim$2\% deviation from the ideal value of 1 in the case of EDR+M2 and up to $\sim$7.5\% deviation in the case of EDR. This difference is due to the worse estimation of the pipeline-reported variance for the first part of the SV program (see \citealt{edr1d} for details). For $\varlss$, its estimation is similar for both samples, showing a clear dependence on wavelength (and hence redshift) for the Lyman-$\alpha$ region and an expected (due to the lack of Lyman-$\alpha$ fluctuations) nearly zero value for the \ion{C}{iii} region. The fit fails at the higher wavelength values in the case of the EDR sample due to the small number of pixels available at these wavelengths, and then the initial value is kept.

Figure \ref{fig:delta_stack} shows the mean flux-transmitted field for the same two regions and samples. In this case, differences between the two samples are smaller, only showing a higher variance in the case of the smaller (EDR) sample. The variances are particularly worse at the largest wavelengths in the Lyman-$\alpha$ region due to the reduced number of pixels available. The \ion{C}{iii} region, for which no re-calibration has been performed, still shows the smooth features already discussed in Section \ref{sec:calibration}. They do not appear in the already re-calibrated Lyman-$\alpha$ region, although there is a higher variance caused by the less number of pixels (as shown in Figure \ref{fig:n_pixels}).

\subsubsection{Example of quasar fit}
\begin{figure*}
	\centering
	\includegraphics[width=1\textwidth]{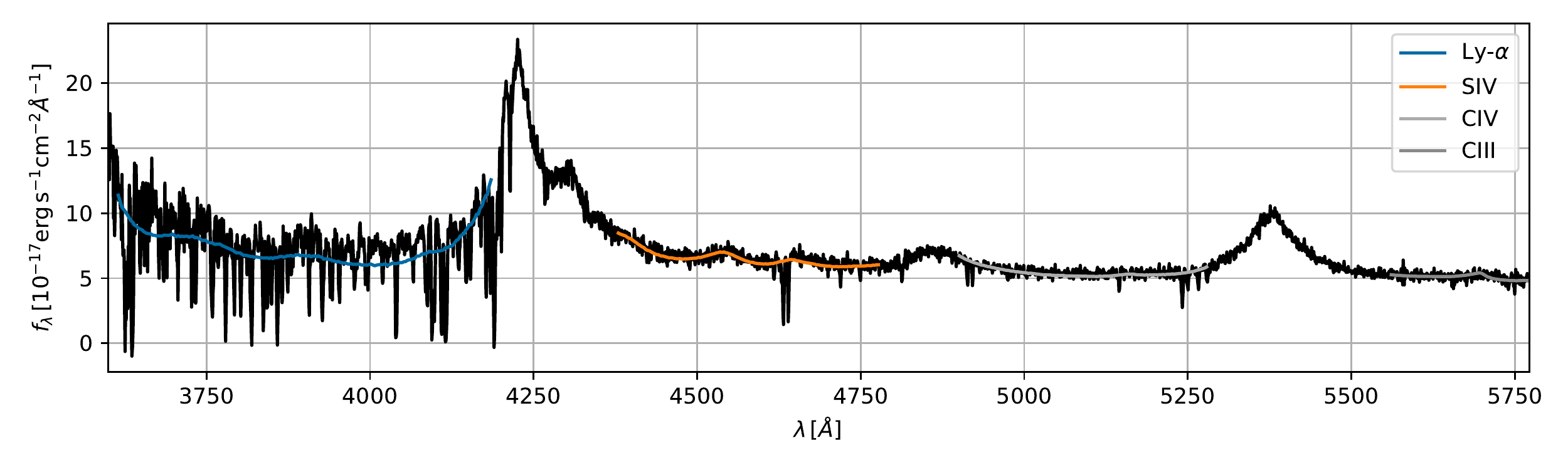}
	\caption{Example of a high signal-to-noise quasar spectrum. The mean expected flux $\overline{C}(\lambda_{\rm RF})$ as in Eq. \ref{eq:meanexpectedflux} is shown for the Lyman-$\alpha$, \ion{S}{iv}, \ion{C}{iv} and \ion{C}{iii} regions. The quasar has a redshift $z_q = 2.495$ and is identified as a DESI object with TARGETID=39628443918272474. As in this example, we occasionally observe metal absorption in the calibration regions.}
	\label{fig:forestexample}
\end{figure*}
An example of a high SNR quasar spectrum is shown in Figure \ref{fig:forestexample}. We can see  the \lyaf at the left side of the Lyman-$\alpha$ emission line ($\lambdarf=1215.7$ \r{A}), with the characteristic absorption features. The solid lines show the mean expected flux for that particular forest at different spectral regions. Metal absorption features can be observed in this example for the \ion{S}{iv} and \ion{C}{iv} re-calibration regions.

\section{Discussion}
\label{sec:discussion}
The main science driver of the Lyman-$\alpha$ catalog presented in this publication is the measurement of 3D correlations that allows us to constraint the BAO scale \citep{edrmain}. 
In this section, we discuss two methodological novelties in the construction of the Lyman-$\alpha$ catalog with respect to previous eBOSS analyses \citep{dMdB_2020}, and we do this by looking at the impact of these changes to the precision with which we will be able to measure the 3D correlations. In particular, in Section \ref{subsec:optima_weights} we discuss the optimization of the weights assigned to each Lyman-$\alpha$ fluctuation, while in Section \ref{subsec:rest_frame_range_choice} we will chosse the rest-frame wavelength range based on the precision of the correlation function measurements.

We conclude this section in \ref{subsec:per_quasar_params} with a discussion on the distribution of per-quasar parameters that capture the diversity of quasar continua in the dataset.

\subsection{Optimal weights}
\label{subsec:optima_weights}

Correlations in the \lyaf are estimated as weighted
averages of products of fluctuations $\delta_q(\lambda)$ in pixel pairs
at a given separation. 
An optimal quadratic estimator would use the inverse of the pixel covariance
as the weight matrix%
, but the large
number of pixels in current Lyman-$\alpha$ datasets makes this inversion 
not feasible. 

Ignoring the small correlation between pixels in different quasar spectra,
one can approximate the covariance as block-diagonal, and make the inversion
tractable. 
This approximation has been used in several measurements of the 1D power
spectrum \citep{McDonald_2006,Karacayli_2020,Karacayli_2022,edr1doptimal},
it has been proposed to measure the 3D power spectrum 
\citep{Font-Ribera_2018}
and it was used in one of the first BAO measurements with the Lyman-$\alpha$
forest \citep{Slosar_2013}.

Recent measurements of 3D correlations in the \lyaf
\citep{Delubac_2015,Bautista_2017,Agathe_2019,dMdB_2020},
on the other hand, have ignored all correlations between pixels and have 
used instead a diagonal weight matrix.
These studies weighted each pixel with the inverse of its variance,
including the instrumental noise and the intrinsic fluctuations (see
Equation \ref{eq:variance_scheme}), effectively approximating the inverse
covariance matrix with the inverse of its diagonal elements.
This approximation is simple and easy to implement, but it is not
the optimal diagonal weight matrix.

In this section we study a simple modification of our diagonal weight matrix,
where we add an extra free parameter $\sigma^2_{\rm mod}$ that modulates the
contribution of the intrinsic fluctuations $\sigma_{\rm LSS}$ to the weights:
\begin{equation}
  w_{q} (\lambda) = \frac{1}{\eta(\lambda) \tilde{\sigma}^2_{\rm pip, q} (\lambda) + \sigma^2_{\rm mod}\varlss (\lambda)} ~.
  \label{eq:weight_scheme_alt}
\end{equation}
A small value of $\sigma^2_{\rm mod}$ is equivalent to weighting the pixels
based solely on the noise variance, while a large value gives the same
weight to all pixels at a given redshift, regardless of instrumental noise.

\begin{figure}
	\centering
	\includegraphics[width=1\columnwidth]{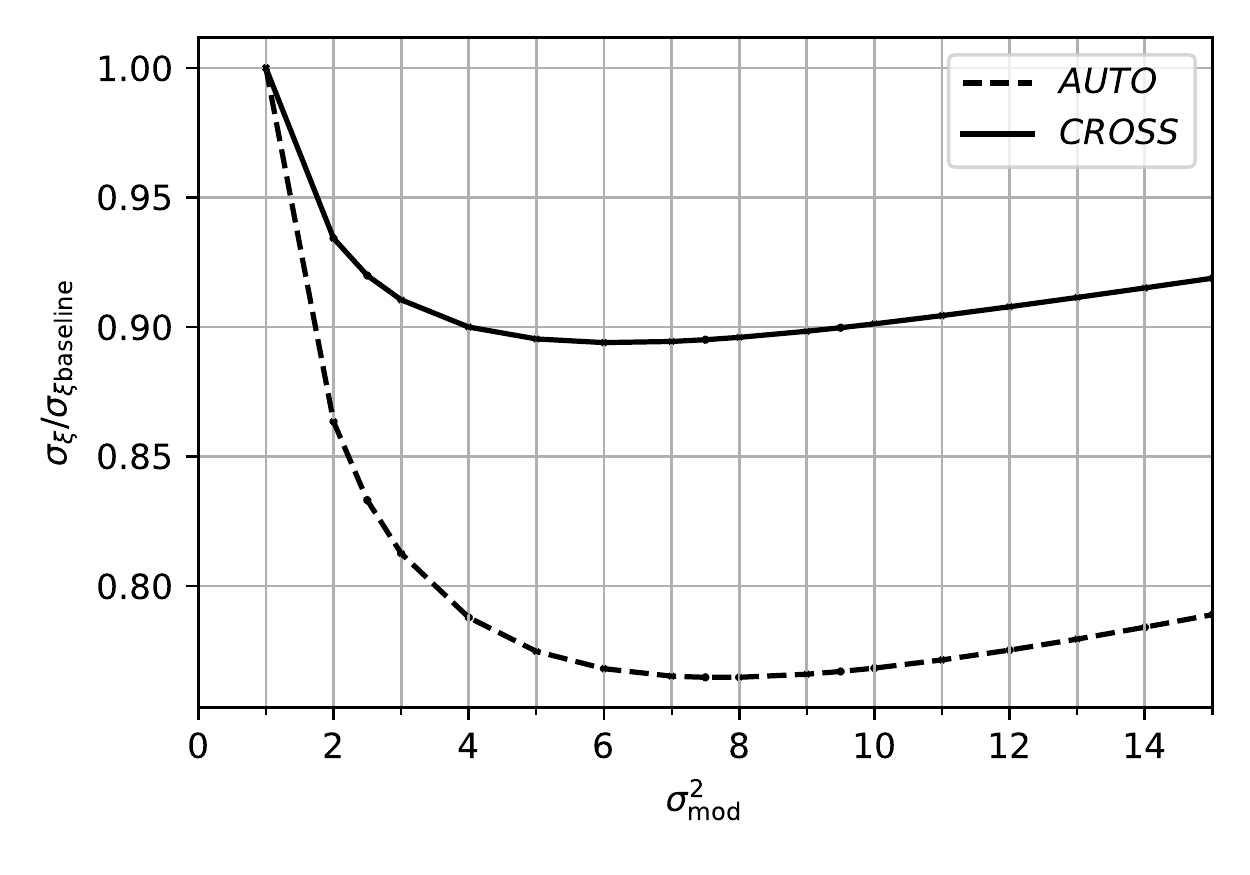}
	\caption{
		Measurement of the Lyman-$\alpha$ auto- (solid) and cross- (dashed) correlation errorbars for different choices of the $\sigma_{\rm mod}^2$ parameter, scaled to a reference value at $\sigma_{\rm mod}^2=1$. The value of the errorbars was averaged over all scales, given the small scale dependency. 
		The $\sigma_{\rm mod}^2$ parameter modified the inverse-variance weighting scheme as defined in Eq. \ref{eq:weight_scheme_alt}, the use of a $\sigma_{\rm mod}^2 \neq 1$ allows us for the optimization of the weighting scheme. In both auto- and cross-correlations, we observe a reduction in the size of the errorbars when we approach the optimal value of the $\sigma_{\rm mod}^2$ parameter. This optimal value is slightly different, but in both cases is found around 7-8. In the optimal value, the improvement in the measurement of the auto-correlation is about 20\%, and 10\% for the case of the cross-correlation. Both measurements have been performed with the full EDR+M2 dataset.
	}
	\label{fig:sigma_mod_effects}
\end{figure}

In Figure \ref{fig:sigma_mod_effects} we show that a value of 
$\sigma_{\rm mod}^2$ around 7-8 can improve the precision of the
auto-correlation measurement by 25\%, at no additional cost.
As expected, the gain in precision in the cross-correlation is roughly half
of that, and we find a 10\% improvement for values of $\sigma_{\rm mod}^2$
around 6-7. 

It is important to note that the actual gain will depend on the properties
of the dataset.
For instance, $\sigma_{\rm pip}$ and $\sigma_{\rm LSS}$ change differently
with the width of the pixels used, and we have tested that when using pixels
of 2.4\AA\ (similar to the ones used in the BAO measurement of 
\citealt{dMdB_2020}) the optimal value is smaller,
$\sigma_{\rm mod}^2 = 3.1$, and the gain in the auto-correlation is only 8\%.

For this publication and the associated Value Added Catalog, we decide to use a value of $\sigma_{\rm mod}^2 = 7.5$, and we leave for future work the implementation of a block-diagonal weighting.

\subsection{Selection of rest-frame wavelength range}
\label{subsec:rest_frame_range_choice}
Due to diversity in quasar spectra near the emission lines, the rest-frame wavelength range that can be used for analyzing the \lyaf is limited. This happens at the red end of the spectra due to the Lyman-$\alpha$ emission line ($\lambdarf = 1215.67$ \r{A}) and at the blue end limited due to the Lyman-$\beta$ line ($\lambdarf = 1025.72$ \r{A}).

Extending the analysis towards longer wavelengths closer to the Lyman-$\alpha$ emission line allows for the incorporation of more data. By including these extra pixels in our analysis, we can improve our measurements of the correlation function if the improvement due to the large number of pixels is not counterbalanced by of the larger pixel variance.

\begin{figure}
	\centering
	\includegraphics[width=1\columnwidth]{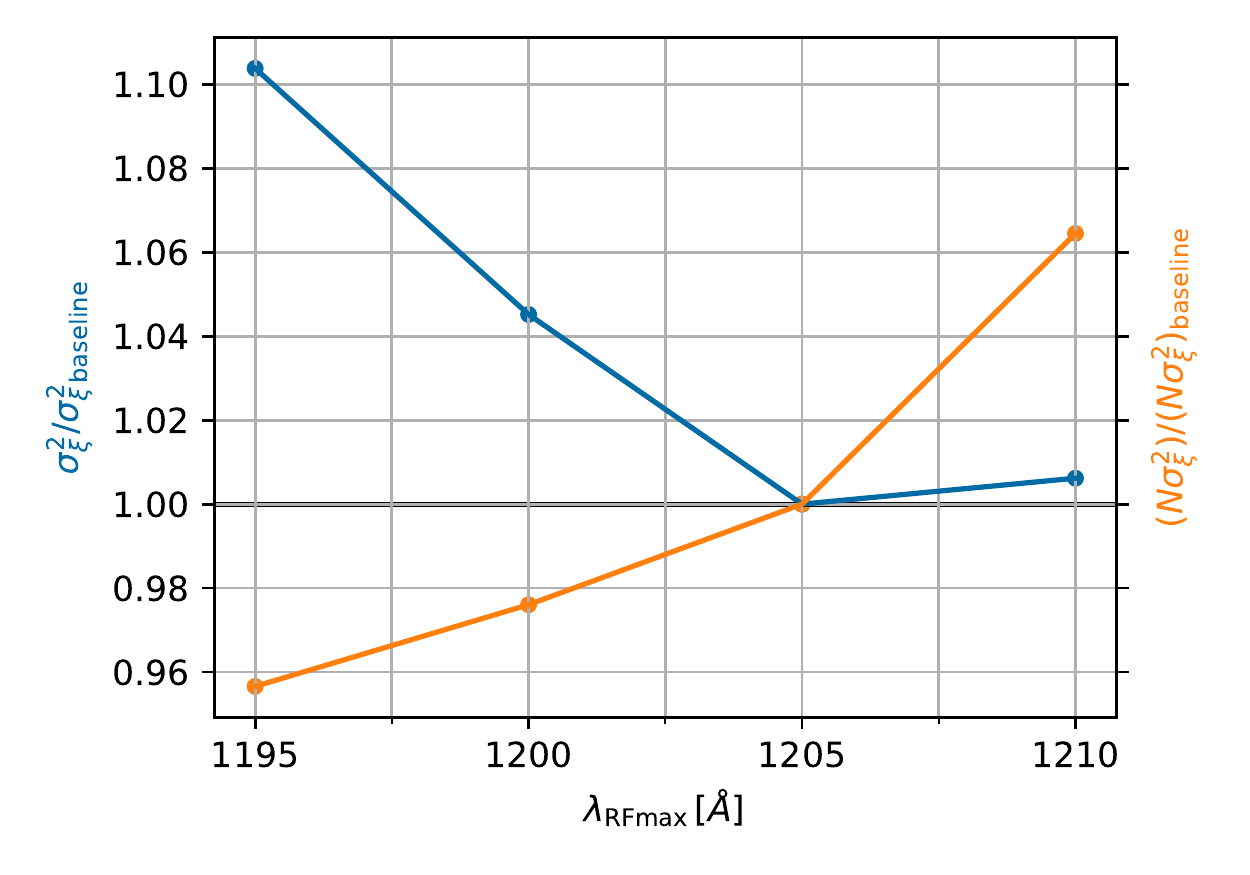}
	\caption{
		Comparison of the size of the errorbars in the Lyman-$\alpha$ auto-correlation for different choices of $\lambda_{\rm RF, max}$. The blue points show the errorbars size after averaging over all scales (given the small scale dependence of this quantity), and scaled to a reference value at $\lambda_{\rm RF, max}=1205$ \r{A}. The orange points show the equivalent measurement but in this case for the product of errorbar sizes times the number of pixel pairs, removing the dependence on the number of pairs used in the measurement. Following the blue points, we observe an improvement in the precision of our auto-correlation measurements when increasing the $\lambda_{\rm RF, max}$ parameter, having the optimal value at 1205 \r{A}. The orange points show that the quality of these added points actually decrease for higher wavelengths, revealing that the improved performance is only driven by the inclusion of more information in our sample. We selected 1205 \r{A} as our default value as a compromise between these two features. We used data from the whole EDR+M2 dataset for these measurements.
	}
	\label{fig:RF_max_errorbars}
\end{figure}

In Figure \ref{fig:RF_max_errorbars}, the blue points show the size of the errorbars in the auto-correlation measurement when we extend the analysis to higher $\lambda_{\rm RF, max}$. The value at $\lambda_{\rm RF, max}=1205$ \r{A} yields the smallest errorbars. To separate the two effects of increased number of pixels and increased pixel variance, we also plot $\sigma^2_{\xi}N$, where $N$ is the number of pairs in the correlation measurement, since we expect $\sigma^2 \propto 1/N$. This will show how valuable the added points are when extending the wavelength range. Its evolution in Figure \ref{fig:RF_max_errorbars} shows that extending $\lambda_{\rm RF, max}$ to higher wavelengths adds less valuable information, and therefore the decrease in the size of the errorbars is only driven by the increase in the sample size.

Given this result, we decided to set $\lambdarf = 1205$ \r{A} because we found that further increasing the limit did not add constraining power. The increased variance near the emission line is likely the reason behind this, as it is not accounted for in our calculations and could potentially affect our measurements if we approached it too closely. A more detailed study of quasar continuum variance, using a larger dataset than the one presented here, is necessary to fully understand its effects. We leave this for future releases of DESI data.

\begin{figure}
	\centering
	\includegraphics[width=1\columnwidth]{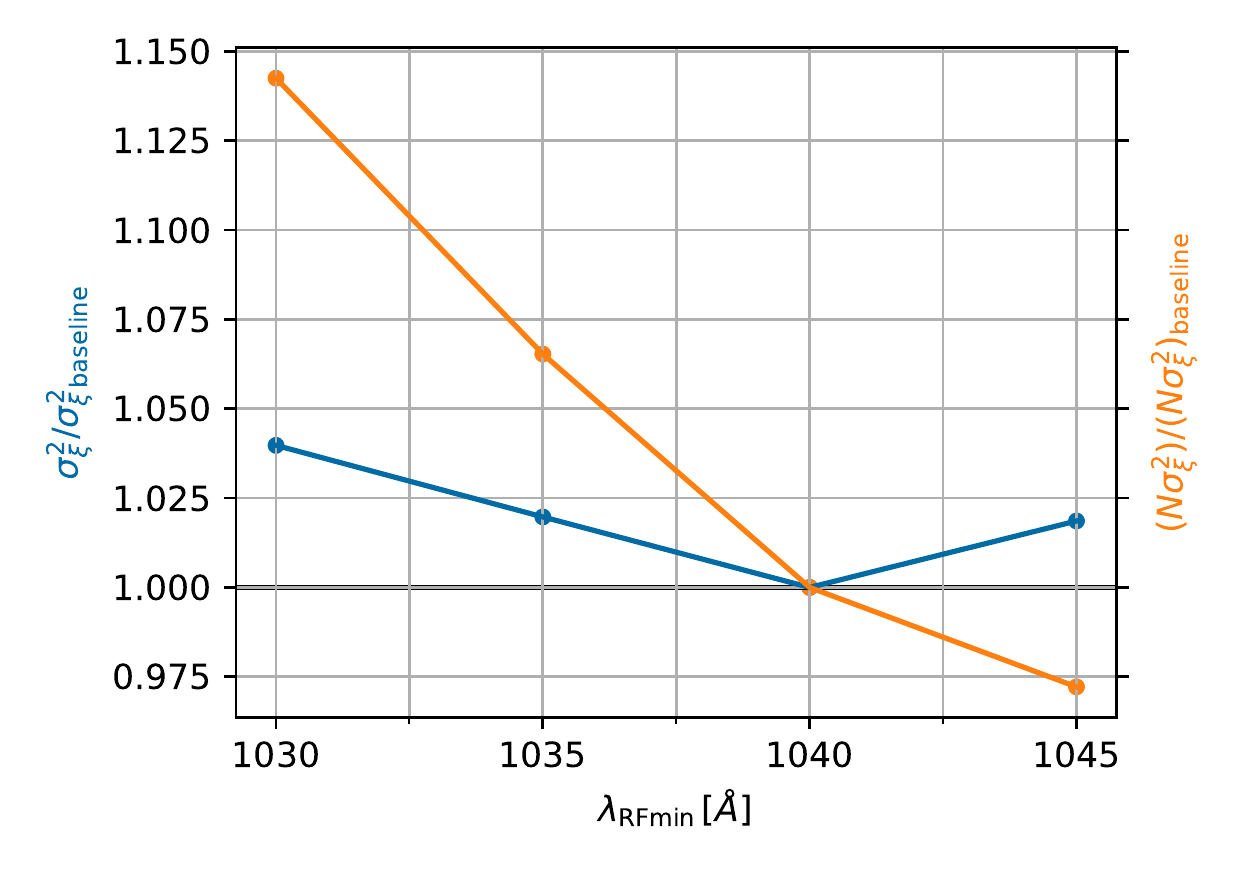}
	\caption{
		Identical measurements as the ones displayed in Figure \ref{fig:RF_max_errorbars}, in this case for values of $\lambda_{\rm RF, min}$. The optimal wavelength is found at $\lambda_{\rm RF, min}=1040$ \r{A}, while having similar features for the orange points: a decrease in the quality of the points added when approaching the emission line. Following the same arguments as in the case of $\lambda_{\rm RF, max}$, we decided to keep the limit at 1040 \r{A}. Again, these measurements were performed using the EDR+M2 sample.
	}
	\label{fig:RF_min_errorbars}
\end{figure}

The same exercise was performed for the blue side of the quasar spectrum, and the results are shown in Figure \ref{fig:RF_min_errorbars}. When compared to the prescription of previous analyses ($\lambda_{\rm RF, min} = 1040$ \r{A} in \citealt{dMdB_2020}), we observe a degradation of the errorbars either going to smaller $\lambda_{\rm RF, min}$ due to the decrease of pixel count; or to larger $\lambda_{\rm RF, min}$ approaching the emission line. For this reason we retain this prescription of $\lambda_{\rm RF, min}= 1040$ \r{A}.

\subsection{Per-quasar parameters (a,b)}
\label{subsec:per_quasar_params}
\begin{figure}
	\centering
	\includegraphics[width=1\columnwidth]{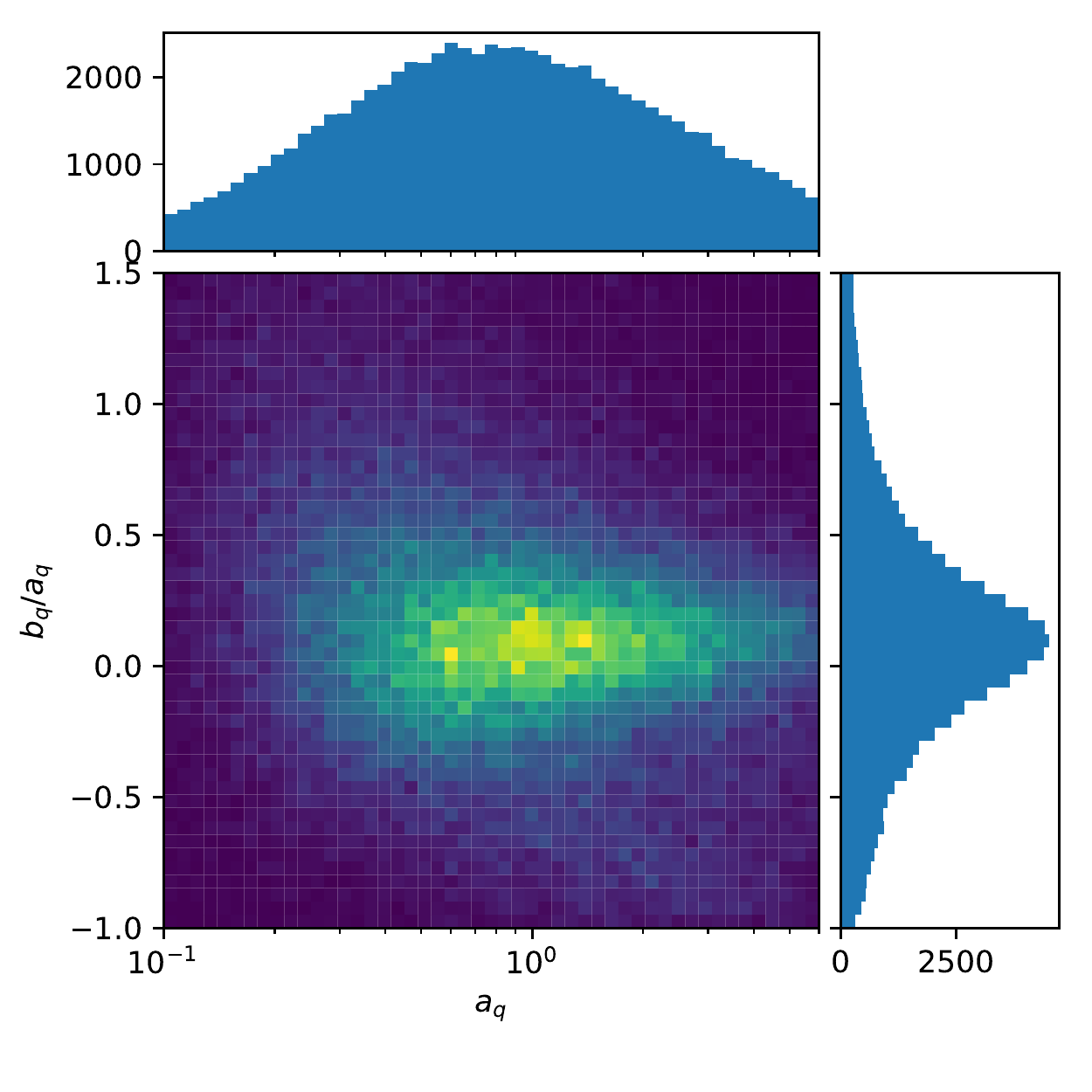}
	\caption{
		Distribution of $a_q$ and $b_q/a_q$ parameters defined in Eq. \ref{eq:meanexpectedflux} for the EDR+M2 sample. The $a_q$ parameter modifies the amplitude of the mean continuum $\overline{C}(\lambdarf)$, while the $b_q/a_q$ term introduces a modification that tilts it as a function of rest-frame wavelength. This last parameter relates to the intrinsic width of the spectral index. The faint long tail at large values of $|b_q/a_q|$ is caused by short forests in the sample, where the continuum fit can be problematic.}
  \label{fig:aqbqhist2d}
\end{figure}
As mentioned in Section \ref{subsec:continuumfitting}, the quasar continua is assumed to follow a universal function of rest-frame wavelength, with a correction by a first degree polynomial parametrized by $a_q$ and $b_q$. Quasar variability can be larger at both ends of the Lyman-$\alpha$ forest, due to the Lyman-$\beta$ and Lyman-$\alpha$ emission lines, but variability in the weak quasar emission lines at $1017 \AA$ and $1123 \AA$ \citep{suzuki2006} could also impact the results. 

In Figure \ref{fig:aqbqhist2d}, we examine the distribution of these two parameters to test this assumption. The plot does not show special features apart from the long tail at large values of $|b_q/a_q|$. This feature is caused by spectra where only a small part of the forest appears in the spectrograph. In these cases, fitting the real continuum of the quasar is difficult, leading to poor fits. For this reason, we require forests to have a minimum length of 150 pixels of 0.8 $\AA$ pixels.

\section{Summary and conclusions}
\label{sec:summary}
In this publication, we present the first measurement of Lyman-$\alpha$ fluctuations using DESI data. We use two data samples: EDR, which mainly consists of SV data; and EDR+M2, which also includes the first two months of the main survey (M2) data. The EDR sample contains 68,750 quasars, of which 20,281 have valid forests for Lyman-$\alpha$ studies, i.e., with the Lyman-$\alpha$ region visible and identifiable. The EDR+M2 sample contains 318,691 quasars, of which 88,511 have valid forests. We release the Lyman-$\alpha$ fluctuations catalog for the EDR sample as part of the first DESI data release.

To achieve this measurement, we adapted the methodology proposed in previous analyses, especially the one outlined in \cite{dMdB_2020}, to suit the unique characteristics of the DESI early data sample. We discussed and applied several important modifications.
\begin{itemize}
	\item We adjusted our pipeline to work with linear-spaced bins in wavelength to match the DESI scheme, preserving its format and precision.
	\item We adapted the re-calibration process by simplifying it, removing unneeded steps that did not improve the performance of the whole re-calibration. We also switched the re-calibration region to allow for more pixels to be used in this process. 
	\item We simplified the weighting scheme by removing terms that added unnecessary complexity and did not contribute to the accuracy of the results, specially for this small data release.
	\item By optimizing our diagonal weight matrix, we improved measurements of the auto-correlation by about  20\%, and of the cross-correlation by about 10\%
\end{itemize}

The presented flux-transmission field catalog could be used for other studies apart from the standard BAO analysis. As in \cite{Font-Ribera_2012,ignasidla,ignasidla2}, its cross-correlation with DLAs can be measured. A similar analysis could be performed with Strong Blended Lyman-$\alpha$ (SBLA) absorption systems, as in \cite{ignasisbla}. The full-shape analysis of the \lyaf three-dimensional correlation function would allow for the measurement of the Alcock-Paczyński effect, as performed in \cite{Cuceu:2023} using eBOSS DR16 data. Furthermore, IGM tomography is also possible with this sample.

As datasets get larger and larger, we will need to be more careful with the analysis. Weights used to measure clustering could be improved in two ways. As discussed in Section \ref{subsec:optima_weights}, the implementation of a block-diagonal weighting scheme will account for correlations between pixels within the same forest. Alternatively, taking into account that quasar diversity makes the estimation of the continua more difficult, one possible way of improving the weights is adding to the weighting scheme proposed in Eq. \ref{eq:variance_scheme} an extra term accounting for errors in the estimation in the continuum. This term would be dependent on $\lambdarf$ and would be larger around the emission lines, where quasar diversity is expected to be higher. Finally, the analysis could be extended into the Lyman-$\beta$ region.

This study can serve as a solid foundation for future research using DESI data. We are excited about the potential for further investigations in this area, as more comprehensive analyses become possible with the availability of future DESI releases.

\section*{Acknowledgments}
CRP is partially supported by the Spanish Ministry of Science and Innovation (MICINN) under grants PGC-2018-094773-B-C31 and SEV-2016-0588. IFAE is partially funded by the CERCA program of the Generalitat de Catalunya. IPR is partially founded by the European Union’s Horizon 2020 research and innovation programme under the Marie Skłodowska-Curie grant agreement No. 754510. AFR acknowledges financial support from the Spanish Ministry of Science and Innovation under the Ramon y Cajal program (RYC-2018-025210) and the PGC2021-123012NB-C41 project, and from the European Union's Horizon Europe research and innovation programme (COSMO-LYA, grant agreement 101044612). IFAE is partially funded by the CERCA program of the Generalitat de Catalunya.

This material is based upon work supported by the U.S. Department of Energy (DOE), Office of Science, Office of High-Energy Physics, under Contract No. DE–AC02–05CH11231, and by the National Energy Research Scientific Computing Center, a DOE Office of Science User Facility under the same contract. Additional support for DESI was provided by the U.S. National Science Foundation (NSF), Division of Astronomical Sciences under Contract No. AST-0950945 to the NSF’s National Optical-Infrared Astronomy Research Laboratory; the Science and Technologies Facilities Council of the United Kingdom; the Gordon and Betty Moore Foundation; the Heising-Simons Foundation; the French Alternative Energies and Atomic Energy Commission (CEA); the National Council of Science and Technology of Mexico (CONACYT); the Ministry of Science and Innovation of Spain (MICINN), and by the DESI Member Institutions: \url{https://www.desi.lbl.gov/collaborating-institutions}. Any opinions, findings, and conclusions or recommendations expressed in this material are those of the author(s) and do not necessarily reflect the views of the U. S. National Science Foundation, the U. S. Department of Energy, or any of the listed funding agencies.

The authors are honored to be permitted to conduct scientific research on Iolkam Du’ag (Kitt Peak), a mountain with particular significance to the Tohono O’odham Nation.

\section*{Data availability}
The \lyaf catalog for the EDR dataset is available in the DESI public database, and can be accessed at \href{https://data.desi.lbl.gov/public/edr/vac/edr/lya/fuji/v0.3}{https://data.desi.lbl.gov/public/edr/vac/edr/lya/fuji/v0.3}. For the case of the EDR+ dataset, it will be released alongside DESI Y1 data. Data points for the Figures in this publication are available in the zenodo repository, at \href{https://doi.org/10.5281/zenodo.8009535}{https://doi.org/10.5281/zenodo.8009535}.

\bibliographystyle{mnras}
\bibliography{bibliography}

\appendix

\section{Changes with respect to previous analyses}
\label{ap:dmdb20procedure}
In this publication, we have presented some changes compared to the last \lyaf analysis conducted using SDSS data \citep{dMdB_2020}. These changes were necessary due to the differences in the way data is structured in DESI as compared to SDSS. One of the most significant is the linear pixelization of the wavelength grid. 

There are three relevant changes in the way the analysis is performed. First, the re-calibration has been updated to account for the improvements offered by the new instrument. Second, the variance estimation process has been simplified. Third, the re-calibration region has been switched to \ion{C}{iii}. We have already discussed the third change in Section \ref{sec:calibration}, now we will describe the first two changes in detail.

\subsection{Re-calibration of spectra in SDSS}
The improved estimation of pipeline noise ($\sigma_{\rm pip}$) in DESI has eliminated the need for multiple re-calibration steps as it was performed in the SDSS analysis.

For SDSS analyses, re-calibration was performed in two steps, both of which were conducted in the same re-calibration region. The first step was equivalent to the one used in this analysis (Section \ref{sec:calibration}). After the first step, a second re-calibration step was added to further correct the reported variance of the pipeline. In this second step, the continuum fitting process was run again to obtain a new estimation of $\eta$.

Then, this new value of $\eta$ is applied in the main Lyman-$\alpha$ fluctuations run to correct for the values of $\sigma_{\rm pip}$:
\begin{equation}
	\sigma_{\rm pip}^{2 \quad ({\rm Ly}\alpha)} (\lambda) = \eta^{({\rm calib2})}(\lambda) \ \sigma_{\rm pip}^{2 \quad ({\rm calib2})} (\lambda).
\end{equation}

The extra re-calibration step in SDSS aimed to make the final estimation of $\eta$ closer to 1, but it also added an unnecessary layer of complexity to the process with the sole purpose of doing so. However, it did not modify the product $\eta \cdot \sigma_{\rm pip,q}^2$, and as a result, the overall variance assigned to pipeline noise was unchanged.

\subsection{Variance estimator}
\label{app:variance_estimator}
In Eq \ref{eq:variance_scheme}, we have defined the model for our variance of the flux $\sigma_q^2(\lambda)$. Its expression has been simplified from previous SDSS analysis, where this variance was modeled as:
\begin{equation}
	\frac{\sigma^2_q(\lambda)}{\left( \overline{F}C_q(\lambda) \right)^2} = \eta(\lambda) \tilde{\sigma}^2_{\rm pip, q} (\lambda) + \varlss (\lambda) + \frac{\epsilon(\lambda)}{\tilde{\sigma}_{\rm pip,q}^2 (\lambda)}.
	\label{eq:variance_scheme_sdss}
\end{equation}
The additional term was added to account for the observed increase in variance at high SNR, which is likely caused by the diversity of quasar spectra.

\begin{figure}
	\centering
	\includegraphics[width=1\columnwidth]{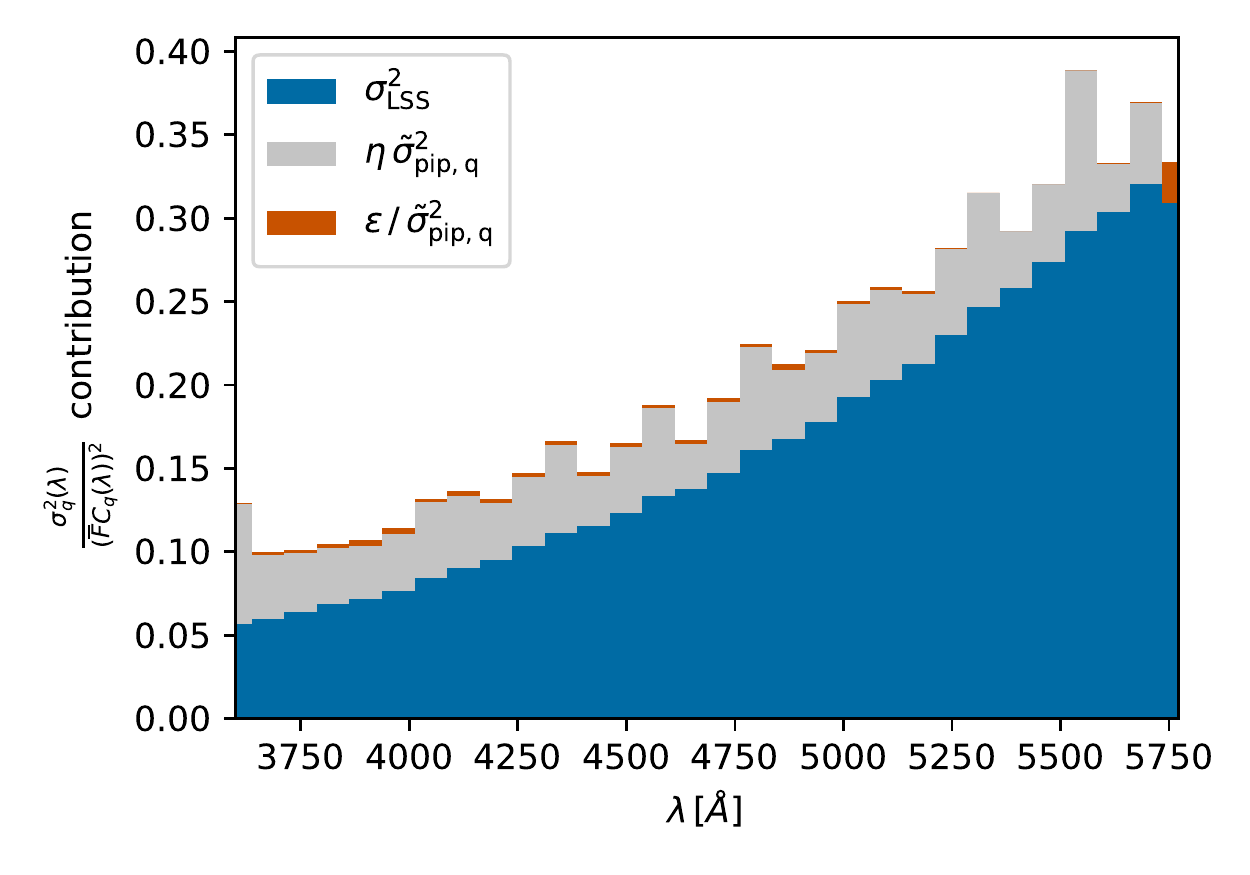}
	\caption{
		Contribution of each term in Eq. \ref{eq:variance_scheme_sdss} to the full variance, for multiple wavelength bins. For this measurement we only used the 5\% of quasars with the highest SNR in the EDR+M2 sample. This analysis reveals that the effect of the $\epsilon$ term in Eq. \ref{eq:variance_scheme_sdss} is minimal, even for the data subset where it is supposed to be most significant.
	} 
	\label{fig:variance_contribution_plot}
\end{figure}

In Figure \ref{fig:variance_contribution_plot}, we present the weight of each term in Eq. \ref{eq:variance_scheme_sdss} contributing to the total variance, selecting only the highest SNR quasars. This shows that the effect of this added term is small, and for simplicity we decided to not include it in our analysis. The plot displays only the top 5 \% higher SNR quasars since this subset is expected to have the largest contribution (the third term in Eq. \ref{eq:variance_scheme_sdss} becomes larger). It is worth noting that the contribution is too small to be discernible when all the objects in the sample are considered.

However, it is important to take into account quasar diversity in future analyses. We plan to incorporate a term in Eq. \ref{eq:variance_scheme} that considers the variance in the quasar's continuum. This term will be dependent on rest-frame wavelength, $\sigma^2_{C} (\lambdarf)$, and will also help us select the region of the quasar suitable for Lyman-$\alpha$ analyses, showing us how close the emission lines we can get.

Nevertheless, for this DESI EDR analysis, we decided to omit this modification since its inclusion lacked sufficient justification, and its influence on the analysis is negligible.

\onecolumn
\section{Catalog description}
The catalog is released in multiple compressed FITS files, one file per healpix pixel. Some additional files, with statistical information of the sample and of each of the continuum fitting steps are also included. Finally the \textsc{picca} configuration file generated at the moment of building the catalog is also provided, this will allow for the reproduction of the catalog for any user having access to DESI data.

Each of the delta files with Lyman-$\alpha$ fluctuations are stored under \texttt{Deltas/} and have the following internal structure:
\begin{verbatim}
	Filename: Delta/delta-16.fits.gz
	No.  Name      Ver  Type      	  Cards  Dimensions  Format
	0    PRIMARY   1	PrimaryHDU    6      ()
	1    LAMBDA    1 	ImageHDU      13   	 (2376,)   	 float64
	2    METADATA  1 	BinTableHDU   33   	 54R x 9C    [K, D, D, D, D, K, 12A, 12A, 12A]
	3    DELTA     1 	ImageHDU      12   	 (2376, 54)  float64
	4    WEIGHT    1 	ImageHDU      11   	 (2376, 54)  float64
	5    CONT      1 	ImageHDU      11   	 (2376, 54)  float64
\end{verbatim}

The \texttt{METADATA} card contains metadata information for each of the forests that belong to the given healpix pixel:
\begin{verbatim}
	XTENSION= 'BINTABLE'           / binary table extension
	BITPIX  =                    8 / 8-bit bytes
	NAXIS   =                    2 / 2-dimensional binary table
	NAXIS1  =                   84 / width of table in bytes
	NAXIS2  =                  315 / number of forests
	PCOUNT  =                    0 / size of special data area
	GCOUNT  =                    1 / one data group (required keyword)
	TFIELDS =                    9 / number of fields in each row
	TTYPE1  = 'LOS_ID  '           / picca line-of-sight id
	TFORM1  = 'K       '           / data format of field: 8-byte INTEGER
	TTYPE2  = 'RA      '           / right ascension
	TFORM2  = 'D       '           / data format of field: 8-byte DOUBLE
	TUNIT2  = 'rad     '           / physical unit of field
	TTYPE3  = 'DEC     '           / declination
	TFORM3  = 'D       '           / data format of field: 8-byte DOUBLE
	TUNIT3  = 'rad     '           / physical unit of field
	TTYPE4  = 'Z       '           / redshift
	TFORM4  = 'D       '           / data format of field: 8-byte DOUBLE
	TTYPE5  = 'MEANSNR '           / mean signal-to-noise ratio
	TFORM5  = 'D       '           / data format of field: 8-byte DOUBLE
	TTYPE6  = 'TARGETID'           / object identification
	TFORM6  = 'K       '           / data format of field: 8-byte INTEGER
	TTYPE7  = 'NIGHT   '           / observation night(s)
	TFORM7  = '12A     '           / data format of field: ASCII Character
	TTYPE8  = 'PETAL   '           / observation petal(s)
	TFORM8  = '12A     '           / data format of field: ASCII Character
	TTYPE9  = 'TILE    '           / observation tile(s)
	TFORM9  = '12A     '           / data format of field: ASCII Character
	EXTNAME = 'METADATA'           / name of this binary table extension
	BLINDING= 'none   '           / blinding scheme used
	COMMENT Per-forest metadata
	CHECKSUM= 'Q93lR70iQ70iQ70i'   / HDU checksum updated 2023-05-20T00:19:40
	DATASUM = '1813998756'         / data unit checksum updated 2023-05-20T00:19:40
\end{verbatim}

The rest of information inside the file comes in ImageHDU format. In order to read it correctly, one has to use the card \texttt{LAMBDA} as observed wavelength. The rest of features: \texttt{DELTA}, \texttt{WEIGHT} and \texttt{CONT} consist of a 2-dimensional array of the values for delta of matter fluctuations, weight assigned to each delta and quasar expected flux. The first index in this 2-dimensional array designates the quasar, and the second the wavelength (the ones given in the \texttt{LAMBDA} image).

\

The auxiliary files are stored under the \texttt{Log/} directory. The file \texttt{delta\_attributes.fits.gz} contains statistical properties of the sample and fit values. Its structure is the following:
\begin{verbatim}
	Filename: Log/delta_attributes.fits.gz
	No.  Name          Ver  Type      	  Cards  Dimensions  Format
	0  	 PRIMARY       1    PrimaryHDU    6   	 ()
	1  	 STACK_DELTAS  1    BinTableHDU   20   	 2376R x 3C  [D, D, D]
	2    VAR_FUNC      1    BinTableHDU   25   	 20R x 6C    [D, D, D, D, J, L]
	3    CONT          1    BinTableHDU   20   	 200R x 3C   [D, D, D]
	4    FIT_METADATA  1    BinTableHDU   24   	 60R x 6C    [K, D, D, D, K, L]
\end{verbatim}
The \texttt{STACK\_DELTAS} card contains the measurement of $\overline{1 + \delta_q(\lambda)}$. \texttt{VAR\_FUNC} stores the fitted variance function, alongside information from the fits. \texttt{CONT} includes the mean expected flux. Finally, the card \texttt{FIT\_METADATA} holds information of the quasar continuum fit for each of the quasars.

Finally, the file \texttt{Log/rejection\_log.fits.gz} includes a single card with metadata for all the objects in the quasar catalog (even the ones rejected) alongside the rejection status.

The full data documentation can be found within the catalog files. 

\section*{Affiliations}
$^{1}$ Institut de F\'{i}sica d’Altes Energies (IFAE), The Barcelona Institute of Science and Technology, Campus UAB, 08193 Bellaterra Barcelona, Spain\\
$^{2}$ Departament de F\'isica, EEBE, Universitat Polit\`ecnica de Catalunya, c/Eduard Maristany 10, 08930 Barcelona, Spain\\
$^{3}$ Department of Physics and Astronomy, University College London, Gower Street, London, WC1E 6BT, UK\\
$^{4}$ IRFU, CEA, Universit\'{e} Paris-Saclay, F-91191 Gif-sur-Yvette, France\\
$^{5}$ Aix Marseille Univ, CNRS/IN2P3, CPPM, F-13288, Marseille, France\\
$^{6}$ Departamento de F\'{i}sica, Universidad de Guanajuato - DCI, C.P. 37150, Leon, Guanajuato, M\'{e}xico\\
$^{7}$ Department of Astronomy and Astrophysics, UCO/Lick Observatory, University of California, 1156 High Street, Santa Cruz, CA 95064, USA\\
$^{8}$ Department of Astronomy and Astrophysics, University of California, Santa Cruz, 1156 High Street, Santa Cruz, CA 95065, USA\\
$^{9}$ Department of Astronomy, Tsinghua University, 30 Shuangqing Road, Haidian District, Beijing, China, 100190\\
$^{10}$ Lawrence Berkeley National Laboratory, 1 Cyclotron Road, Berkeley, CA 94720, USA\\
$^{11}$ Center for Cosmology and AstroParticle Physics, The Ohio State University, 191 West Woodruff Avenue, Columbus, OH 43210, USA\\
$^{12}$ Department of Physics, The Ohio State University, 191 West Woodruff Avenue, Columbus, OH 43210, USA\\
$^{13}$ The Ohio State University, Columbus, 43210 OH, USA\\
$^{14}$ Instituto de F\'{\i}sica Te\'{o}rica (IFT) UAM/CSIC, Universidad Aut\'{o}noma de Madrid, Cantoblanco, E-28049, Madrid, Spain\\
$^{15}$ Consejo Nacional de Ciencia y Tecnolog\'{\i}a, Av. Insurgentes Sur 1582. Colonia Cr\'{e}dito Constructor, Del. Benito Ju\'{a}rez C.P. 03940, M\'{e}xico D.F. M\'{e}xico\\
$^{16}$ Kavli Institute for Cosmology, University of Cambridge, Madingley Road, Cambridge CB3 0HA, UK\\
$^{17}$ Department of Physics, The University of Texas at Dallas, Richardson, TX 75080, USA\\
$^{18}$ Department of Astronomy, The Ohio State University, 4055 McPherson Laboratory, 140 W 18th Avenue, Columbus, OH 43210, USA\\
$^{19}$ Astrophysics Group, Department of Physics, Imperial College London, Prince Consort Rd, London, SW7 2AZ, UK\\
$^{20}$ Department of Physics, University of Warwick, Gibbet Hill Road, Coventry, CV4 7AL, UK\\
$^{21}$ Department of Physics \& Astronomy, University  of Wyoming, 1000 E. University, Dept.~3905, Laramie, WY 82071, USA\\
$^{22}$ Instituto Avanzado de Cosmolog\'{\i}a A.~C., San Marcos 11 - Atenas 202. Magdalena Contreras, 10720. Ciudad de M\'{e}xico, M\'{e}xico\\
$^{23}$ Aix Marseille Univ, CNRS, CNES, LAM, F-13388, Marseille, France\\
$^{24}$ Instituto de Astrof\'{i}sica de Canarias, C/ Vía L\'{a}ctea, s/n, E-38205 La Laguna, Tenerife, Spain \\  Universidad de La Laguna, Dept. de Astrof\'{\i}sica, E-38206 La Laguna, Tenerife, Spain\\
$^{25}$ Sorbonne Universit\'{e}, CNRS/IN2P3, Laboratoire de Physique Nucl\'{e}aire et de Hautes Energies (LPNHE), FR-75005 Paris, France\\
$^{26}$ Excellence Cluster ORIGINS, Boltzmannstrasse 2, D-85748 Garching, Germany\\
$^{27}$ University Observatory, Faculty of Physics, Ludwig-Maximilians-Universit\"{a}t, Scheinerstr. 1, 81677 M\"{u}nchen, Germany\\
$^{28}$ Beihang University, Beijing 100191, China\\
$^{29}$ Physics Dept., Boston University, 590 Commonwealth Avenue, Boston, MA 02215, USA\\
$^{30}$ Department of Physics \& Astronomy, University College London, Gower Street, London, WC1E 6BT, UK\\
$^{31}$ Department of Physics and Astronomy, The University of Utah, 115 South 1400 East, Salt Lake City, UT 84112, USA\\
$^{32}$ Instituto de F\'{\i}sica, Universidad Nacional Aut\'{o}noma de M\'{e}xico,  Cd. de M\'{e}xico  C.P. 04510,  M\'{e}xico\\
$^{33}$ Department of Physics, Southern Methodist University, 3215 Daniel Avenue, Dallas, TX 75275, USA\\
$^{34}$ Departamento de F\'isica, Universidad de los Andes, Cra. 1 No. 18A-10, Edificio Ip, CP 111711, Bogot\'a, Colombia\\
$^{35}$ Observatorio Astron\'omico, Universidad de los Andes, Cra. 1 No. 18A-10, Edificio H, CP 111711 Bogot\'a, Colombia\\
$^{36}$ NSF's NOIRLab, 950 N. Cherry Ave., Tucson, AZ 85719, USA\\
$^{37}$ Instituci\'{o} Catalana de Recerca i Estudis Avan\c{c}ats, Passeig de Llu\'{\i}s Companys, 23, 08010 Barcelona, Spain\\
$^{38}$ Department of Physics and Astronomy, Siena College, 515 Loudon Road, Loudonville, NY 12211, USA\\
$^{39}$ Department of Physics and Astronomy, University of Sussex, Brighton BN1 9QH, U.K\\
$^{40}$ National Astronomical Observatories, Chinese Academy of Sciences, A20 Datun Rd., Chaoyang District, Beijing, 100012, P.R. China\\
$^{41}$ Department of Physics and Astronomy, University of Waterloo, 200 University Ave W, Waterloo, ON N2L 3G1, Canada\\
$^{42}$ Perimeter Institute for Theoretical Physics, 31 Caroline St. North, Waterloo, ON N2L 2Y5, Canada\\
$^{43}$ Waterloo Centre for Astrophysics, University of Waterloo, 200 University Ave W, Waterloo, ON N2L 3G1, Canada\\
$^{44}$ Department of Physics and Astronomy, Sejong University, Seoul, 143-747, Korea\\
$^{45}$ CIEMAT, Avenida Complutense 40, E-28040 Madrid, Spain\\
$^{46}$ Space Telescope Science Institute, 3700 San Martin Drive, Baltimore, MD 21218, USA\\
$^{47}$ Department of Physics \& Astronomy, Ohio University, Athens, OH 45701, USA\\
$^{48}$ University of Michigan, Ann Arbor, MI 48109, USA\\
\bsp
\label{lastpage}
\end{document}